\documentclass[9pt,twocolumn,twoside]{pnas-new}
\usepackage{graphics,color,graphicx,amsmath}
\usepackage{xr}

\templatetype{pnasresearcharticle} 

\newcommand{\br}[1]{\left<#1\right>}

\newcommand{\dn}[0]{\dot{n}}
\newcommand{\dr}[0]{\dot{r}}
\newcommand{\Dr}[0]{\Delta r}
\newcommand{\rmax}[0]{r_{\rm max}}
\newcommand{\DQ}[0]{\Delta Q}
\newcommand{\rmin}[0]{r_{\rm min}}
\newcommand{\aroot}{\alpha_{\rm r}}
\newcommand{\croot}{c_{\rm r}}

\title{Dynamics of growth, death, and resource competition in sessile organisms}

\author[a]{Edward D.~Lee}
\author[a]{Christopher P.~Kempes} 
\author[a]{Geoffrey B.~West}

\affil[a]{Santa Fe Institute, 1399 Hyde Park Rd, Santa Fe, NM 87501}
\dates{This manuscript was compiled on \today}

\leadauthor{Lee} 

\significancestatement{Though fairy circles and termite mounds stand out as examples of remarkably regular patterns emerging over long times from local interactions, ecological spatial patterns range from regular to random and temporal patterns from transient to stable. We propose a minimal quantitative framework to unify this variety by accounting for how quickly sessile organisms grow and die mediated by competition for fluctuating resources. We reproduce a wide range of spatial patterns and predict transient features such as population shock waves that align with demographic measurements. Our work may help, by connecting diverse ecological dynamics, apply lessons from model systems more broadly including to inferring local ecological conditions via remote mapping.}

\authorcontributions{All authors helped develop the initial idea. EDL did the analysis, wrote the code, and initially drafted the manuscript. All authors helped edit the manuscript.}
\authordeclaration{The authors declare no competing interests.}
\correspondingauthor{\textsuperscript{2}To whom correspondence should be addressed. E-mail: edlee@santafe.edu}

\keywords{ecology $|$ population dynamics $|$ spatial patterning $|$ metabolic scaling $|$ competition}

\begin{abstract}
Population-level scaling in ecological systems arises from individual growth and death with competitive constraints. We build on a minimal dynamical model of metabolic growth where the tension between individual growth and mortality determines population size distribution. We include resource competition based on shared capture area separately. By varying relative rates of growth, death, and competitive attrition, we connect regular and random spatial patterns across sessile organisms from forests to ants, termites, and fairy circles. Then, we consider transient temporal dynamics in the context of asymmetric competition that primarily weakens the smaller of two competitors such as canopy shading or large colony dominance. When such competition couples slow timescales of growth with fast competitive death, it generates population shock waves similar to those observed in forest demographic data. Our minimal quantitative theory unifies spatiotemporal patterns across sessile organisms through local competition mediated by the laws of metabolic growth which in turn result from long-term evolutionary dynamics.
\end{abstract}

\begin{document}

\maketitle
\thispagestyle{firststyle}
\ifthenelse{\boolean{shortarticle}}{\ifthenelse{\boolean{singlecolumn}}{\abscontentformatted}{\abscontent}}{}

\begin{figure*}[tb]\centering
	\hspace{-2.5cm}\includegraphics[width=.75\linewidth]{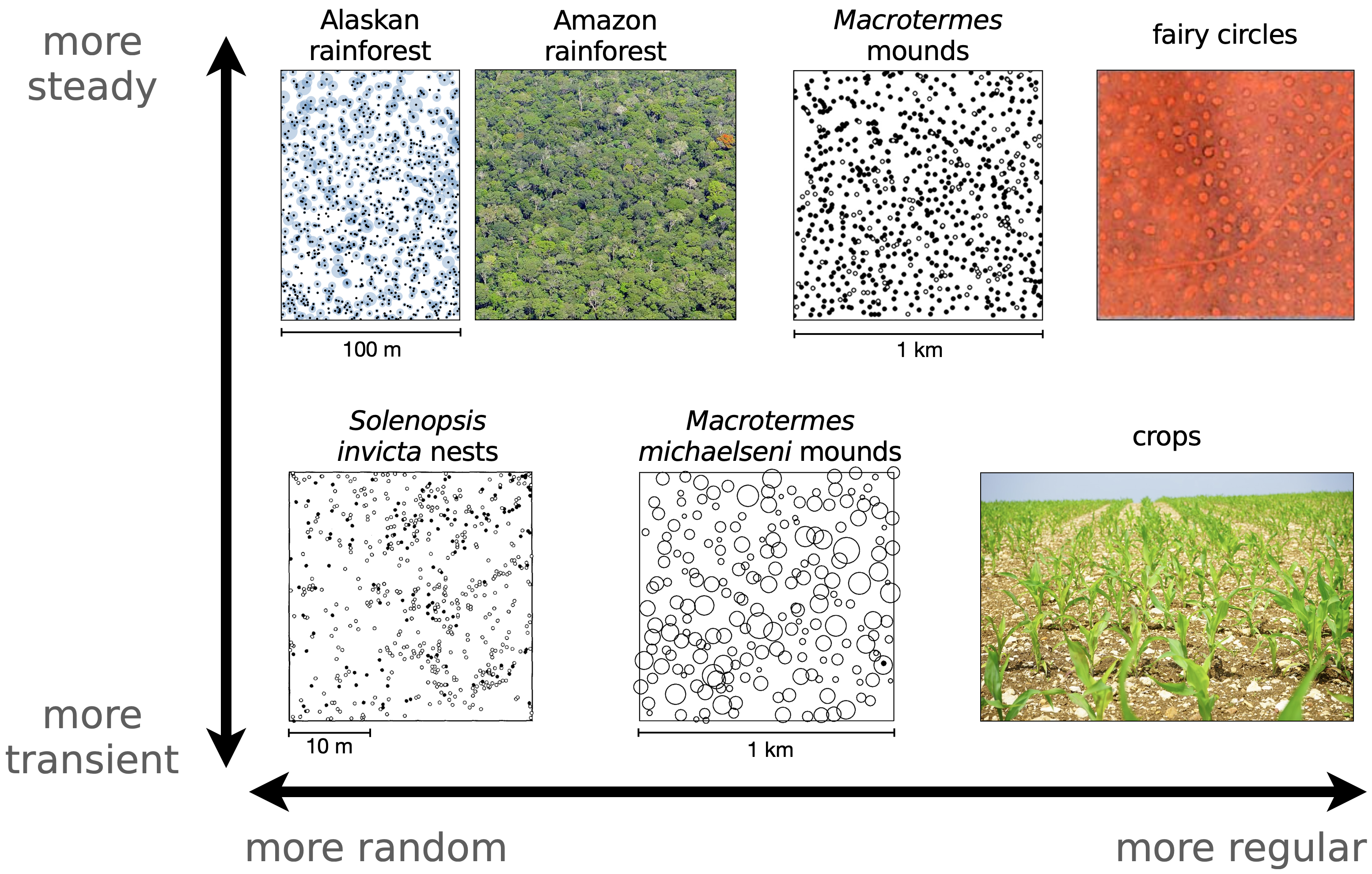}
	\caption{Regular to random spatial distributions and transient to slow temporal evolution in sessile organisms. (top row) Trees in Alaskan rainforest (circles indicate basal stem diameter of $>2.5$\,cm increased by a factor of 5) \cite{schneiderSoutheastAlaska2020}, view of the Amazon canopy \cite{palmerAerialView2011}, semi-regularly packed termite mounds reprinted from reference \cite{muvengwiGeologyDrives2018} (empty circles are inactive mounds), and hexagonally packed fairy circles reprinted from \cite{tschinkelLifeCycle2012}. (bottom row) Newly built ant nests (reprinted by permission from Springer Nature: Springer {\it Oecologia}, ``Spatial dynamics of colony interactions in young populations of the fire ant Solenopsis invicta,'' Adams \& Tschinkel, 1995) \cite{adamsSpatialDynamics1995}, termite mounds with size shown by circles (reprinted by permission from Springer Nature: Springer {\it Insectes Sociaux}, ``Multi-scale pattern analysis of a mound-building termite species,'' Grohmann et al., 2010) \cite{grohmannMultiscalePattern2010}, and perennial agricultural crops. Dynamics range from transience-dominated, in the case of crops razed at the end of the season or newly built ant nests that die within days as indicated by open circles, to long-lasting structures such as fairy circles which can live individually for decades or forests at demographic equilibrium lasting millenia. Scale is unavailable for fairy circles, but they range from 2 to 12\,m in diameter meaning that the shown plot covers some hundreds of meters on a side \cite{tschinkelLifeCycle2012}. 
	}\label{gr:organism examples}
\end{figure*}

\dropcap{E}cological niches display a wide variety of spatial and temporal patterns ranging from random to regular and from transient to long-lived. In Figure~\ref{gr:organism examples}, we show a small sample from such diversity including the remarkable fairy circles in semi-arid environments \cite{tschinkelLifeCycle2012}, regular and random tiling of termite mounds \cite{grohmannMultiscalePattern2010,muvengwiGeologyDrives2018}, and more randomly spaced ant nests and trees \cite{adamsSpatialDynamics1995,schneiderSoutheastAlaska2020}. This variation is not limited to between taxa but also varies between different plots in the same region. These systems also operate on different timescales, where fairy circles have estimated lifetimes of around half of a century compared to days or weeks for nascent ant nests and centuries for trees in unperturbed forests. In the extreme, transient growth is maximized for agricultural crops which are then razed at maturity before demographic stability \cite{dengModelsTests2012,westobySelfThinningRule1984}. Overall, fast and slow dynamics of sessile organisms are characterized by a range of spatial distributions, from the random to the regular, that reflect underlying forces of growth, death, and competition.   

The mechanisms underlying such pattern formation have been a source of robust debate especially in the context of vegetation  \cite{pringleSpatialSelfOrganization2017,rietkerkSelfOrganizedPatchiness2004}. Following Turing's seminal work on scale-dependent feedback, namely local activation and long-range inhibition, similar principles of pattern formation with local density dependence have been considered \cite{cahnFreeEnergy1958,turingChemicalBasis1990,levinHypothesisOrigin1976,liuPhaseSeparation2016}, touching on the more general question of how multiple scales of time and space emerge \cite{levinProblemPattern1992,rietkerkRegularPattern2008,staverSpatialPatterning2019}. More recent work has connected these principles with mechanisms of biological interaction and environmental feedback \cite{borgognoMathematicalModels2009,tarnitaTheoreticalFoundation2017,farriorTheoryPredicts2019}. 
For spatial patterning, approaches to mechanism range from using perturbations like cascades of tree death to explore self-organized criticality in forests \cite{soleSelfsimilarityRain1995,schefferEarlywarningSignals2009}, to mapping the Turing-like activation-inhibition concepts onto scale-dependent plant processes \cite{rietkerkRegularPattern2008,staverSpatialPatterning2019} which could be modulated by environmental conditions \cite{deblauweEnvironmentalModulation2011}, to ecosystem engineers that by modifying the local environment generate bare and densely vegetated patches \cite{giladEcosystemEngineers2004,tarnitaTheoreticalFoundation2017}. Demographic theories, in contrast, focus on variables that aggregate across species and space such as age and size \cite{lotkaAnalyticalNote1920,vonfoersterRemarksChanging1959,sinkoNewModel1967} and build on allometric dependence of growth, mortality, and resource acquisition \cite{lorimerSimulationEquilibrium1984,kohyamaSimulatingStationary1991,kohyamaTreeSpecies2003,coomesDisturbancesPrevent2003,muller-landauComparingTropical2006,westGeneralQuantitative2009,enquistExtensionsEvaluations2009,linPlantInteractions2013,farriorDominanceSuppressed2016}. In an alternative set of approaches, mechanism-free maximum entropy principles can capture demographic patterns by fixing a few population ``state variables'' to predict measured properties \cite{harteMetabolicPartitioning2017}. Across these examples, forests are particularly well-studied empirically across diverse species, sizes, and environments \cite{hubbellBarroColorado2005,muller-landauTestingMetabolic2006} and grounded on predicted theoretical regularities in space and demography such as in the context of metabolic scaling \cite{westGeneralModel1999,niklasInvariantScaling2001,enquistDoesException2007,enquistInvariantScaling2001} and mechanical limits \cite{savageSizingAllometric2008,jensenPhysicalLimits2013}.

Here, we build on previous work on forest growth to consider sessile organisms broadly in the context of both spatial structure and demographic dynamics. We propose a minimal dynamical model that integrates timescales of individual growth and mortality with competitive attrition on a background of fluctuating resources. With the model, we study the emergence and erasure of spatiotemporal order in ecological systems. We show how competition alone is insufficient to generate strong spatial regularity and that additional constraints on growth and death are essential for spatial order. Since most ecological systems are out of equilibrium, we extend our model to consider transient phenomena and predict population shock waves as a feature of competitive interactions when there is metabolic growth. This minimal framework serves to unify at a conceptual level the role of various timescales for pattern formation in distinct ecological settings.

As the starting point, we consider how metabolism determines individual growth and death. Metabolic scaling theory describes the origins of scaling laws in organism growth across a large range of body sizes derived from energetic constraints \cite{westGeneralModel1997,westGeneralModel1999,kempesPredictingMaximum2011}. Given constraints on average resource consumption per unit area, individual growth follows power law, allometric scaling relations connecting accumulation of biomass $m$ or the organism's physical dimensions such as the stem radius $r$ with age. In the context of forests where individuals are fixed in location, metabolic scaling can be connected with population-level statistics such as spatial density, biomass production, and stand energetics determined by the balance of individual growth and mortality \cite{enquistExtensionsEvaluations2009,enquistInvariantScaling2001}. Such predictions have been verified for individual organisms \cite{niklasInvariantScaling2001,moriMixedpowerScaling2010} and have highlighted ecosystem-level regularities such as total population density and predator-prey relations \cite{damuthInterspecificAllometry1987,hattonPredatorpreyPower2015}. Regularities suggest that unifying principles act across systems \cite{franklinOrganizingPrinciples2020,niklasPlantAllometry2004} such as from energetic constraints \cite{enquistExtensionsEvaluations2009,enquistInvariantScaling2001,enquistAllometricScaling1999,yeakelDynamicsStarvation2018}.

One surprising prediction of metabolic scaling theory is that local competitive interactions are unnecessary to explain population distributions \cite{enquistExtensionsEvaluations2009,westGeneralQuantitative2009}. Yet, local competition is what drives long-term evolutionary dynamics to optimize fundamental energetic constraints \cite{darwin1909origin}. Moreover, competition coupled with other timescales can introduce complex dynamics \cite{durrettSpatialAspects1998,strogatzNonlinearDynamics1994} such as in response to exogenous perturbations \cite{levinDisturbancePatch1974,andereggWhenTree2016}, which goes beyond steady-state assumptions. Other than mechanistic additions to metabolic scaling theory \cite{savageSizingAllometric2008}, competition, perturbation, and other dynamics present potential explanations for significant and sometimes substantial deviations from predictions \cite{muller-landauTestingMetabolic2006,enquistDoesException2007}. Here, we present a minimal model to account for these missing factors. 

We start with allometric scaling theory of forest growth in Section~\ref{sec:compartment model} and connect deviations from metabolic scaling theory to organism density, resource variability, and competitive interactions in Section~\ref{sec:symmetric competition}. We explore the implications of competition through space in Section~\ref{sec:space} and time in Section~\ref{sec:dynamics}, concluding with Section~\ref{sec:discussion}. Though we explicitly rely on the language of forests, referring to individuals as trees and dimension as stem radius, we discuss a general formulation that extends to other sessile organisms.

\section{A size class model for population growth}\label{sec:compartment model}
The fractal structure of a forest exists both at the physical branching of individual trees as well as in the way that populations of larger trees ``branch off'' into smaller ones. This self-similar, fractal structure reflects energetic constraints that shape the long-term evolutionary dynamics of forest life \cite{westGeneralQuantitative2009,moriMixedpowerScaling2010}. Connecting energy expenditure with physical limits of vasculature on which energy is distributed leads to allometric scaling theory of growth. When applied to the rate of basal stem radius growth $\dr$ \cite{westGeneralModel1997,westGeneralModel1999},
\begin{align}
	\dr(r) &\approx \frac{3}{8}c_m^{1-b} \bar{a} r^{b}. \label{eq:allo growth}
\end{align}
For sufficiently long times $r\sim t^{1/(1-b)}$ for time $t$. Eq~\ref{eq:allo growth} expresses the general principle of biomass production in terms of a constant determined from biological energetics $\bar a$, how radius scales with tree mass $m$, $r=c_mm^{3/8}$, and growth exponent $b=1/3$. Other sessile organisms fill available space determined by analogous mechanisms of growth, death, and competition, suggesting that metabolic principles provide a bridge from well-studied forests to sessile organisms more generally \cite{pringleSpatialSelfOrganization2017}.

Building on the metabolic picture of growing individuals, we consider size classes labeled by radial dimension $r_k$ with index $k$ of population number $n_k(t)$ as a function of time $t$. Using forests as our example, these size classes group together trees of various species, roles, and micro-environments, and so we describe properties averaged over such variety. The smallest size class $k=0$ is filled with saplings of stem radius $r_0$ that have grown from seedlings with rate $g_0$ \cite{enquistDoesException2007}. As new saplings appear in the system, older ones grow into the next class $k=1$, reflected in the rate of change of stem radius $\dr_k$, where the discrete classes encompass stems of radius within the interval $[r_k,r_k+\Delta r)$. Furthermore, trees die with a size-dependent inherent mortality rate $\mu_k$, which we consider independent of competition-based mortality. Accounting for these individual properties of metabolic growth and death, we obtain a dynamical equation for change in population number per unit time for saplings,
\begin{align}
	\dn_0(t) &= g_0 - n_0(t) [\dr_0/\Delta r+\mu_0].\label{eq:dn0}
\end{align}
For larger trees, the change in the population is determined by the rate at which smaller plants in size class $k-1$ grow into the size class $k$, 
\begin{align}
	\dn_k &= n_{k-1} \dr_{k-1}/\Delta r - n_k[\dr_k/\Delta r+\mu_k],\label{eq:dnk}
\end{align}
describing a sequence of ever larger tree sizes that are populated by an incoming flux of younger and smaller trees and depopulated as trees grow to a larger size or die. Thus, Eqs~\ref{eq:dn0} and \ref{eq:dnk}, without specifying the particular functional forms for growth $\dr_k$ and mortality $\mu_k$, describe the simplest possible form for independent tree growth without reference to either environment or local competitors.

Though population is typically binned into discrete size classes in both observation and theory, tree growth is in reality a function of continuous radius $r$. Relating the index $k$ to radius $r$ such that $r_k \equiv r_0 + k\,\Delta r$, we obtain
\begin{align}
	\dn(r,t) &= -\partial_r [n(r,t)\dr(r)] - n(r,t)\mu(r) \label{eq:dn1}
\end{align}
with sapling boundary condition
\begin{align}
	\dn(r_0,t) &= g_0 - n(r_0,t)[\dr(r_0)/\Dr+\mu(r_0)]. \label{eq:dn2}
\end{align}
also known as demographic equilibrium theory when referring to steady state \cite{muller-landauComparingTropical2006} (see Appendix~\ref{sec:continuum model}). These equations determine the continuum formulation of the size class model, including only growth and natural mortality as a starting hypothesis.  

Taking predictions from allometric scaling theory that relate mass growth function $dm/dt$ with tree radius in Eq~\ref{eq:allo growth}, we obtain a functional form for mortality \cite{enquistExtensionsEvaluations2009,muller-landauTestingMetabolic2006,clarkScalingPopulation1993}. With Eq~\ref{eq:allo growth} and $n(r)\propto r^{-\alpha}$ and at stationarity $\dn(r,t)=0$,
\begin{align}
\begin{aligned}
	\mu(r) &= \bar{A} r^{b-1},\\
	\bar{A} &= \frac{3}{8}\bar a c_m^{1-b}[\alpha-b].
\end{aligned} 
\end{align}
Thus, stationarity directly fixes the form of metabolic mortality in the simple size class model from Eqs~\ref{eq:dn1} and \ref{eq:dn2} in a way that determines the population number exponent,
\begin{align}
	\alpha &= b + \frac{8\bar{A}}{3\bar a c_m^{1-b}}.\label{eq:alpha}
\end{align}
The population number exponent in Eq~\ref{eq:alpha} indicates the role of metabolic growth in the first term and the relative timescales of growth and death in the second. When growth dominates, we would recover $\alpha\approx b=1/3$ and population number is determined solely by the growth curve, whereas when mortality overtakes individuals quickly, $\alpha\rightarrow\infty$, and no trees survive beyond birth. When metabolic growth is determined by a power law, the simple size class model fixes the forms of scaling in mortality and population as a combination of both the exponent driving growth but also the relative timescales at which mortality and growth act \cite{muller-landauComparingTropical2006}.

From this minimal model of tree growth under the scaling assumptions of individual tree allometry, we obtain a wide range of possible steady states encompassing both predictions of metabolic scaling theory as well as virtually any population number scaling. This reflects the fact that space-filling in forests, when $\alpha=2$, does not depend separately on typical growth and mortality rates but is determined by the ratio of the scaling coefficients, which may be fixed by energetic constraints. Since these features only determine the exponent, deviations from space-filling at steady state, such as for size distributions observed in large trees (Figure 1 in reference \cite{enquistExtensionsEvaluations2009}), could arise from processes such as competitive interactions which are not included in a model only accounting for metabolic scaling.

\section{Competition for fluctuating resources}\label{sec:symmetric competition}
Resource collection in sessile organisms is mediated through local area. Local nutrient collection describes foraging by ants and termites \cite{tarnitaTheoreticalFoundation2017}, where proximate sources of material and energy drive growth while scarcity induces physiological stress and death \cite{yeakelDynamicsStarvation2018}. Likewise trees obtain nutrients, water, and sunlight through overlapping root or canopy volumes \cite{purvesPredictingUnderstanding2008,kempesPredictingMaximum2011}, and competition is largely determined by area overlap between neighbors \cite{farriorTheoryPredicts2019}. Overlapping canopies in particular reduce light available to shorter trees but not to taller ones \cite{farriorDominanceSuppressed2016}, an example of asymmetric competitive interaction that we discuss later. As a general formulation of the consequences of symmetric competitive interactions \cite{adamsSpatialDynamics1995}, we consider how resource availability is modulated by overlap and environmental fluctuations relative to basal metabolic need.

All organisms have some basal resource budget $Q_0$ above which growth is feasible. For local resource capture, we expect the budget to scale with physical dimension to some exponent $\eta_1$ and constant parameter $\beta_1$, or $Q_0(r)=\beta_1 r^{\eta_1}$, inspired by observations for trees \cite{kempesPredictingMaximum2011}.\footnote{In the context of water uptake, $\eta_1\approx1.8$ \cite{kempesPredictingMaximum2011}.} If the total amount of resource per unit area is a time-fluctuating quantity $\rho(t)$, then the amount of resource that tree i could potentially obtain from resource area $a_{\rm i}\equiv a(r_{\rm i}) \propto r^{2\alpha_r}$ is $\rho(t)a_{\rm i}$. Beyond periodic diurnal patterns, long-time averaged resource distribution fluctuates randomly about the mean $\bar\rho$, captured by division with a random variable $\xi$ representing scarcity, $\rho(t) = \bar\rho/\xi(t)$. Noting that in some cases --- such as durations of low precipitation \cite{petersCriticalPhenomena2006}  (Appendix Figure~\ref{gr:si pr rainfall}) --- resource fluctuations can be modeled accurately with power law tails, we consider a scale-free distribution of fluctuations $h(\xi) = \xi_0^{1-\nu}\xi^{-\nu}$, where $\xi_0$ ensures that $\bar\xi=1$.\footnote{When $\nu<2$, we must also fix an upper limit to the distribution to ensure a finite mean.} The exponent $\nu$ primarily summarizes whether extreme events dominate the distribution $1<\nu<2$ or fluctuations are tightly limited $\nu>2$. Though time-averaged resource availability may determine maximum tree size, it is the fluctuations below basal metabolic requirements that induce mortality.

Putting these together, incoming resource rate depends on the amount of sharing that tree i with resource area $a_{\rm i}$ does with neighbor j with overlap in area $\Delta_{\rm i}a_{\rm j}$,
\begin{align}
	\DQ(t) &= \varepsilon\rho(t) \left[a_{\rm i} - f\sum_{\langle\rm i j\rangle}\Delta_{\rm i} a_{\rm j}\right]- Q_0(r_{\rm i}).\label{eq:DQ}
\end{align}
Eq~\ref{eq:DQ} indicates a resource extraction efficiency $\varepsilon$, a sum over the neighbors $\br{\rm ij}$ indexed j of tree i, and a constant fraction $f\in[0,1]$ of resources siphoned off given overlap with each competing neighbor. When $f=1/2$, competitors equally split available resources, a zero-sum game, whereas for $f>1/2$ competition reduces resource availability overall as if individuals pay a cost for competing and for $f<1/2$ resources are reusable or relationships symbiotic. When $\DQ<0$ such as with large overlap or high scarcity, mortality from resource stress occurs with rate $s$ such that trees are sensitive to resource deprivation when $s\gg1$ and relatively robust to such fluctuations when $s\ll1$. Thus, Eq~\ref{eq:DQ} captures the balance of basal metabolic needs with resource competition that strengthens with overlapping area.

\begin{figure}\centering
	\includegraphics[width=\linewidth]{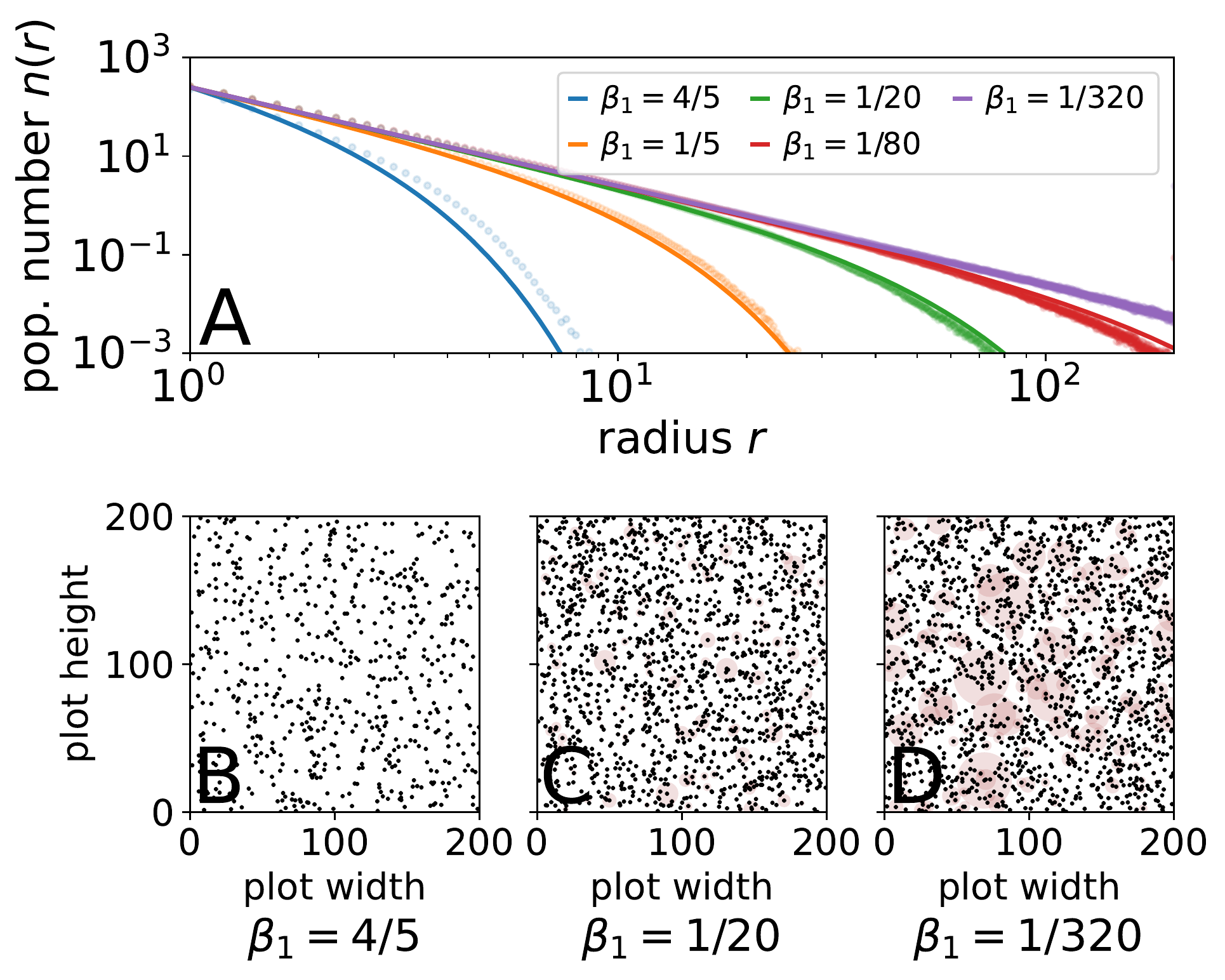}
	\caption{(A) Population number $n(r)$ with varying strength of area competition ($\alpha=2$, $\nu=5/2$, $L=200$, averaged over time and $K=15$ random forests). Mean-field approximation (solid lines) mirrors the shape of the 2D forest simulation (circles) for varying basal metabolic coefficient $\beta_1$. (B--D) Simulated forest plots. Brown circles represent root competition area centered about black dots.}\label{gr:compartment model2}
\end{figure}

Averaging over many spatial arrangements over a long period of time, we consider the mean-field effect from such competition (see Appendix~\ref{si sec:symmetric competition} for details)
\begin{align}
	\Delta\bar{Q}(t) &= \varepsilon\rho(t)a_{\rm i}\left[ 1 - f \right] - Q_0(r_{\rm i}).\label{eq:mft DQ}
\end{align}
Thus, resource competition with neighbors saps fraction $f$ from total incoming resource flux at any given time $\varepsilon \rho(t) a_{\rm i}$. The approximation in Eq~\ref{eq:mft DQ} is an accurate description when considering many trees over a large area that just fill the available space and interact weakly, but it assumes that interactions are stronger than in two dimensions. Then, competitive attrition rate matters when incoming resources are insufficient to cover basal metabolic rate. With probability $p(\xi>\xi_{\rm basal})=\int_{\xi_{\rm basal}}^\infty h(\xi)\,d\xi$ such insufficiency occurs, where $\Delta\bar{Q}=0$ defines a minimum sustainable level of scarcity $\xi_{\rm basal}$. The resulting probability of fatal fluctuations is
\begin{align}
\begin{aligned}
	p(\xi>\xi_{\rm basal}) &= B r^{\kappa},\\
	B = \rmax^{-\kappa}, &\quad \kappa = (\nu-1)(\eta_1-2\alpha_r).
\end{aligned}\label{eq:resource compet}
\end{align}
The probability is normalized by constant $B$ set by recognizing that there is only a single largest tree with radius $r_{\rm max}$ by definition, which then relates the phenomenological parameters $\varepsilon$ and $f$. When resource area grows slower than metabolic need, as is the case for $\eta_1>2\alpha_r$, larger trees have less margin for low resources because they sit close to the boundary of basal metabolic need. Yet if it were possible (though unrealistic) for resource area to grow faster, $\eta_1<2\alpha_r$, then growth is unconstrained and larger trees have more buffer to withstand environmental fluctuations.\footnote{In the marginal case $\eta_1=2\alpha_r$, the coefficient of competition modifies the scaling exponent $\alpha$. This is a mathematical possibility but unrealistic.} Furthermore, the form of exponent $\kappa$ shows that growth in metabolic cost is mediated by the fluctuations in resource availability described by exponent $\nu$ such that when $\nu>2$, its distribution is narrow and we expect there to be sharp difference in the impact of competition for large trees below and above a cutoff. For $\nu\rightarrow1$, all trees, small or large, will pay substantial costs for competition. 

Combining metabolic growth and mortality from Eqs~\ref{eq:dn1} and \ref{eq:dn2} and competition from Eq~\ref{eq:resource compet}, we obtain the generalized size class model
\begin{align}
\begin{aligned}
	\dn(r,t) &= -\partial_r[n(r,t)\,\dr(r)] - n(r,t)[\mu(r) + Bs\,r^{\kappa}].
\end{aligned}\label{eq:general mft}
\end{align}
By solving Eq~\ref{eq:general mft} for steady state, we find
\begin{align}
	n(r) &= \tilde n_0 \left(\frac{r}{r_0}\right)^{-\alpha} \exp\left( -\frac{F}{\kappa+1-b} r^{\kappa+1-b} \right),\label{eq:stationary soln}
\end{align}
with normalization constant $\tilde n_0$ and $F\equiv 8B sc_m^{b-1}/3\bar{a}$. This shows that an exponentially decaying tail truncates the simple scaling form, imposing a cutoff on a scale of $[F/(\kappa+1-b)]^{-1/(\kappa+1-b)}$. Eq~\ref{eq:general mft} is the general formulation incorporating both metabolic scaling theory and the impact of area-mediated resource use from pairwise competitive interactions, yielding the product of a scaling law with a decaying tail that wiggles as resource fluctuations are varied.

We show numerical simulations of an explicit two-dimensional simulation of trees in a large plot in Figure~\ref{gr:compartment model2}A in comparison with the mean-field approximation from Eq~\ref{eq:stationary soln}. The mean-field approximation does not exactly capture the tail of the distribution, but it does surprisingly well. Furthermore, it captures the qualitative intuition that for large trees, $r\gg1$, metabolic constraints dominate, introducing an upper cutoff on the maximum possible tree size in the system whose radius varies with the growth coefficient. The suddenness of this cutoff is controlled by the fluctuation exponent $\nu$ such that we find strong curvature away from purely scale-free metabolic scaling in the stationary distribution for smaller $\nu$. 

Importantly, the mean-field argument clearly links resource area with resource fluctuations. This means that deviations from metabolic scaling theory may result from different combinations of resource area growth and fluctuations in a way that makes effects hard to disentangle \cite{coomesDisturbancesPrevent2003}. Beyond the particular form of competitive interactions we consider, this framework is naturally extendable by, for example, modifying resource sharing fraction $f$ to reflect cooperative or non-cooperative interactions or to change allometrically. Such modifications do not change underlying metabolic scaling but do change the probability of fatal fluctuations, $p(\xi>\xi_{\rm basal})$. Its derived scaling form hints that an infinite variety of competitive effects may be summarized by exponent $\kappa$, a possible indication of universality originating in physical scaling.

\section{From random to regular spatial patterns}\label{sec:space}
Different organisms, and even the same organism in another environment, may show systematic variation in spacing \cite{mujinyaSpatialPatterns2014}. Such variation reflects individual growth dynamics and the strength of resource-area-based competition due to the local properties of competitor species or the environment \cite{levinHypothesisOrigin1976,westobySelfThinningRule1984,slatkinModelCompetition1984}. Returning to Figure~\ref{gr:organism examples}, we again point out randomness in spatial surveys of an Alaskan rainforest along with {\it Macrotermes michaelseni} mounds and ant nests. In contrast, the spacing between {\it Macrotermes falciger} mounds is more regular. Besides inter-taxonomic variation, there is also evidence of random and systematic variation between different plots in nearby regions.\footnote{See Figure~4 in reference \cite{mujinyaSpatialPatterns2014} for termites in several soil types and Figure~\ref{gr:si forest examples} for variation amongst plots in Alaskan rainforest.} Thus, natural spatial distributions of sessile organisms may be attributable to the assorted effects of individual allometries and local competition in our model.

\begin{figure}[t]\centering
	\includegraphics[width=.9\linewidth]{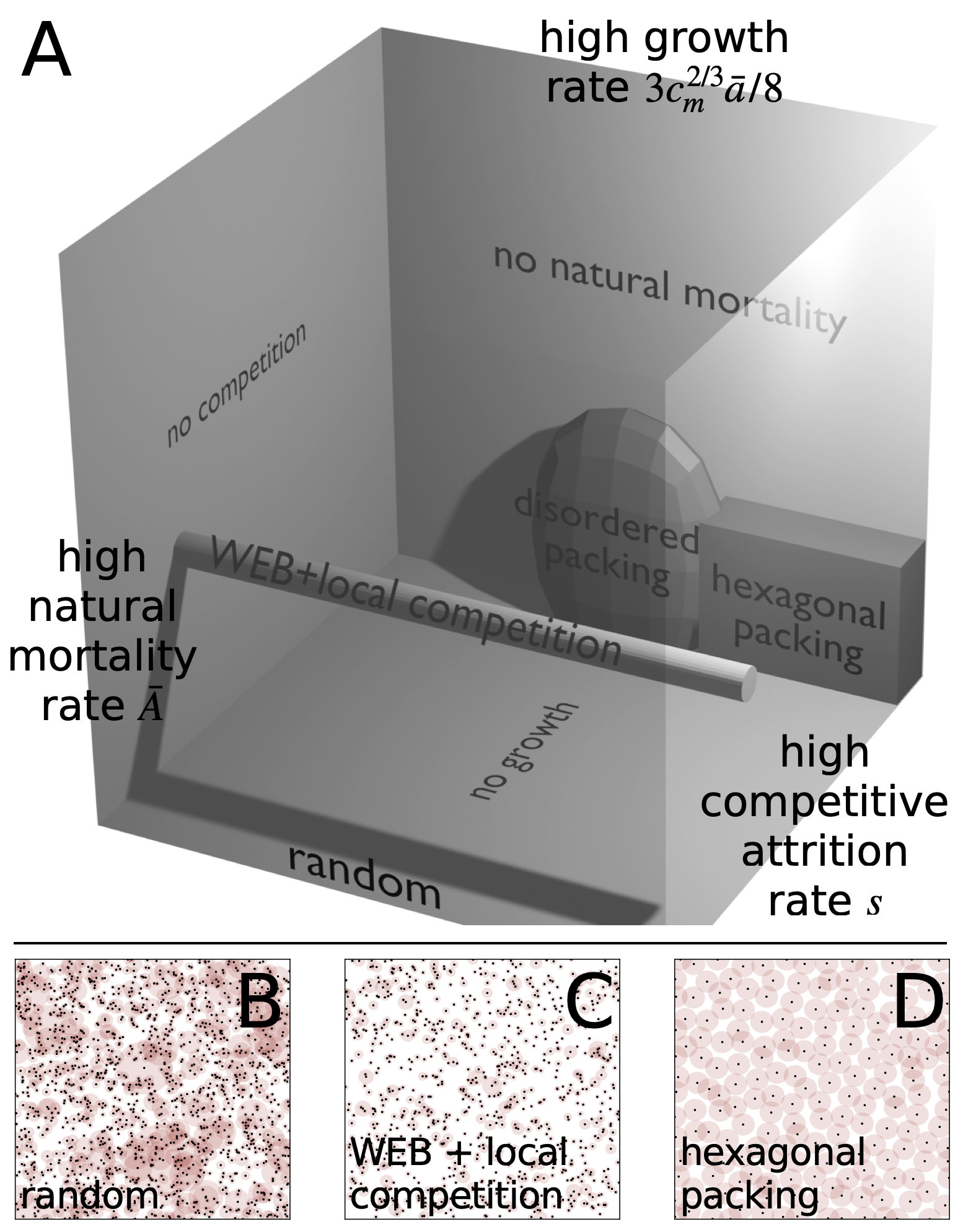}
	\caption{(A) Schematic realm of models defined on rates of growth, death, and competitive attrition. WEB theory of allometric forest growth corresponds to fixing population number exponent $\alpha=2$ while varying competitive attrition rate (cylinder). Regular hexagonal packing only emerges in a tight limit where growth and death rates approach zero and competitive death rate is high. (B--D) By varying timescales, we obtain a variety of spatial patterns qualitatively similar to examples in Figure~\ref{gr:organism examples}.}\label{gr:phase space}
\end{figure}

We survey such variety in Figure~\ref{gr:phase space} along a schematic realm of spatial patterns generated by our model. We traverse this realm by varying the rates of growth, death, and competition in Eq~\ref{eq:general mft}. The planes jutting out from the back corner in Figure~\ref{gr:phase space}A all correspond to theories where one of the terms is negligible. When competitive attrition is negligible, or $s\rightarrow0$, population scaling is pinned to the plane where there is a perfect scaling law determined by mortality and growth. For example, the idealized WEB model for forest distributions contains only growth and death and corresponds to the point where $\alpha = 2$ \cite{enquistExtensionsEvaluations2009,westGeneralQuantitative2009}. The other limits of no natural mortality or no growth lead to qualitatively different configurations that may mimic other spatial patterns found across sessile organisms. In this sense, this realm of models is a three-dimensional slice of a much higher-dimensional set of models with different exponents, as opposed to rates, yet sufficient to capture qualitative variety in ecological spatial patterning.

\begin{figure}[t]\centering
	\includegraphics[width=\linewidth]{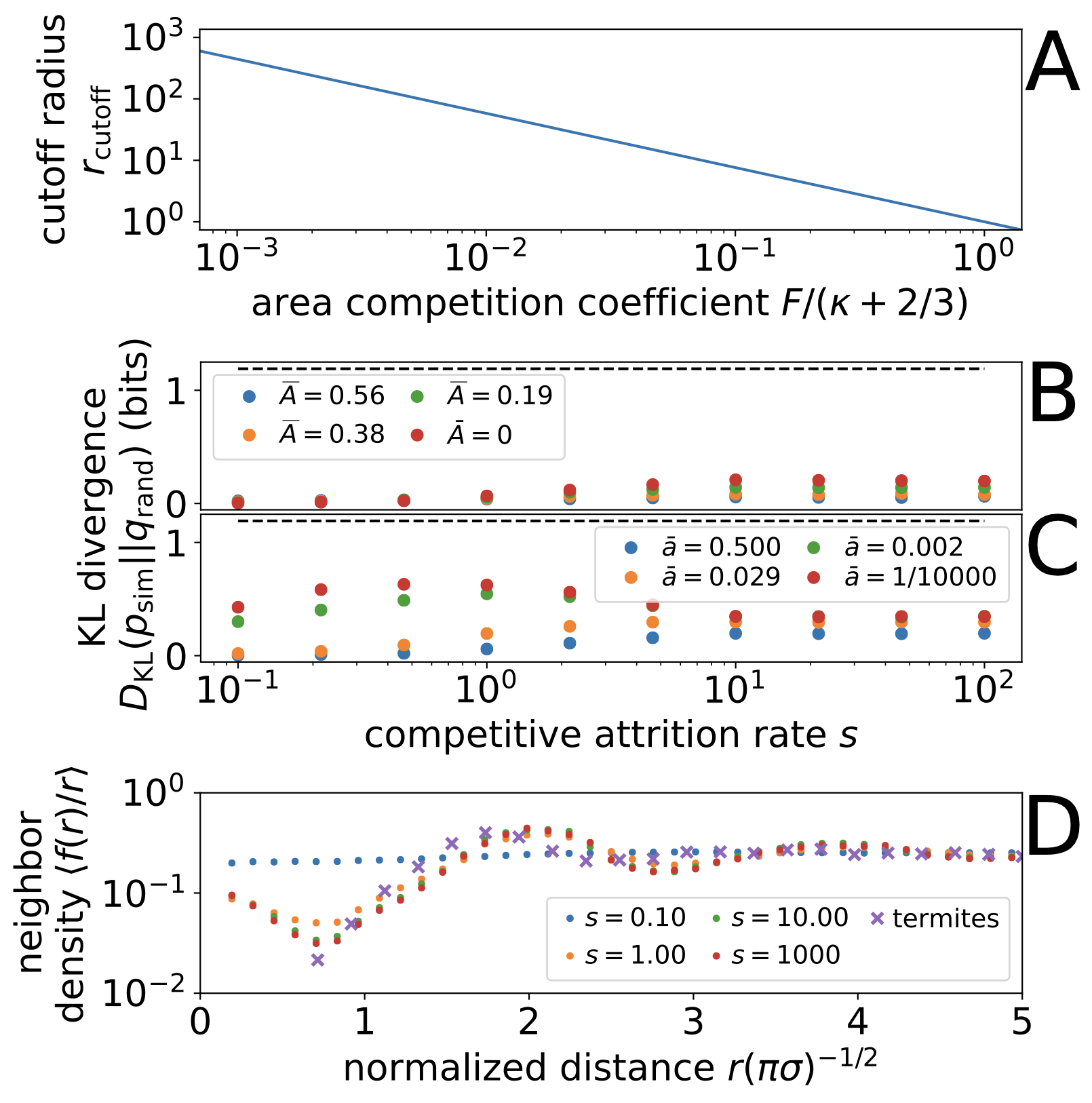}
	\caption{Characterizing trajectories through realm of models. (A) Location of cutoff to population number $n(r)$ decreases as competitive costs increase, $r_{\rm cutoff} \equiv [F/(\kappa+2/3)]^{-1/(\kappa+2/3)}$, tracing the gray cylinder in Figure~\ref{gr:phase space}. (B \& C) Deviations from randomness indicate emergence of order measured by KL divergence of the nearest-neighbor distance distribution for simulation $p(\rmin)$ from random points $q(\rmin)$. Panel B shows randomness-dominated phase when growth rate is significant, $3c_m^{2/3}\bar{a}/8=0.3$ despite small mortality rate $\bar A$ as competitive attrition rate $s$ is driven up. Panel C shows signs of ordering when mortality rate is negligible $\bar A=0$ while growth rate $\bar a$ is driven to zero. Around $s=1$, we find a ``liquid'' phase, where organisms are densely packed but without long-range order. Dashed black line indicates KL divergence measured at the ``solid'' phase. Bin size set to $\Dr=1/20$. (D) Normalized neighbor density at distance $r$, $\br{f(r)/r}$, indicates ``solid,'' hexagonally packed phase. Function $f(r)$ counts all neighbors at distance $r$ with $f(0)=1$ and is plotted against distance normalized by average spacing $1/\sqrt{\pi\sigma}$ given density $\sigma$. For comparison, we show the neighbor density for Namibian termite mounds from reference \cite{tarnitaTheoreticalFoundation2017}, which are more tightly packed than in our simulation.}\label{gr:phase space scan}
\end{figure}

In the limit of weak interaction, the spatial distribution of individuals is random. Then, the probability of not encountering any neighbors within a distance $\rmin$ is given by the Poisson distribution with average $\sigma\pi \rmin^2$, with individual density $\sigma$. However, for finite plots like the ones we consider in Figure~\ref{gr:phase space} and forest plot surveys, it is essential to account for corrections from points sitting near the boundaries. The typical number of points close to the boundaries for a unit square is $\eta = 2\sigma$, and these only have half of the typical number of neighbors. As a result, the probability of the nearest neighbor being at distance greater than $\rmin$ is the mixture
\begin{align}
	q(\rmin) &= (1-\eta) \exp(-\sigma \pi \rmin^2) + \eta \exp(-\sigma \pi \rmin^2 / 2). \label{eq:neighbor distribution}
\end{align}
Competitive interactions manifest as deviations from the prediction of Eq~\ref{eq:neighbor distribution}. As a measure of difference between the random distribution $q(\rmin)$ and observation $p(\rmin)$, we rely on a principled quantitative measure, the Kullback-Leibler divergence \cite{coverElementsInformation2006},
\begin{align}
	D_{\rm KL}[p||q] &= \int_0^\infty p(\rmin)\log_2 \left( \frac{p(\rmin)}{q(\rmin)} \right)\,d\rmin.\label{eq:dkl}
\end{align}
Calculation of Eq~\ref{eq:dkl} requires determining a bin size for integration, as is discussed further in SI Section~\ref{sec:si dkl}. Eq~\ref{eq:dkl} represents a holistic way of measuring the strength of competitive interactions using nearest-neighbor distances in contrast with mean measures like overdispersion that do not account for the shape of the distribution \cite{slatkinModelCompetition1984}.

Moving across the gray cylinder extending out from WEB theory in Figure~\ref{gr:phase space}A corresponds to strengthening competitive interactions by increasing competitive attrition rate $s$. This region describes a set of models with a fixed population scaling exponent but with cutoffs in population scaling changing as in Eq~\ref{eq:stationary soln} and shown in Figure~\ref{gr:phase space scan}A. When sufficiently strong, such effects obscure population scaling. As we show in Figure~\ref{gr:phase space}B, however, strong variation in population number is not reflected in the statistics of nearest-neighbor separation until metabolic processes are severely suppressed. When fixing growth rate $3c_m^{2/3}\bar{a}/8=0.3$ and varying death rate $\bar{A}$, as we do in the top graph, we find that the nearest-neighbor distribution hardly changes. Once we fix $\bar A=0$ and drive growth rate to zero simultaneously as we do in the bottom graph, however, we begin to see the emergence of a different phase. In this limit and for moderate competitive attrition $m$, the system condenses into a disordered packing, liquid-like phase (Figure~\ref{gr:liquid phase}). Nevertheless, long-range order fails to appear because nearest-neighbor statistics are dominated by disorder introduced by turnover from randomly placed seedlings and continuously changing tree size.
In other words, metabolic growth and death act on sufficiently fast timescales that regular patterns in spacing take too long to stabilize. In organisms with different rules for metabolic scaling, we may expect to find stronger tendencies for self-organization.

Such an example manifests in the semi-regular packing of the fairy circles shown in Figure~\ref{gr:organism examples}. Such spacing entails a relatively narrow and peaked distribution of mound areas at some maximum size, a phenomenon incompatible with allometric growth. Instead, this distribution implies that mounds that approach the maximum size are stable and that strong competitive interactions inhibit the formation of new smaller mounds. We can approximate such dynamics by driving growth and natural mortality to zero and vastly enhancing competitive mortality. This ensures that mounds are fixed at a typical size, with rigid boundaries delineated by strong competitive interactions, and close-to-hexagonal spacing that minimizes survival of randomly placed colony seeds.\footnote{In the ``zero-temperature'' limit where competitive mortality always selects out the weaker of two competitors, it is clear that tight packing is stable to disorder. Hexagonal packing is the densest of packings and thus most stable to infiltration.} As we draw in black dashed lines across Figures~\ref{gr:phase space scan}B and C, the simultaneous limits of slow growth ($\bar{a}\rightarrow0$), slow death ($\bar{A}\rightarrow0$), and lethal competition ($s\rightarrow\infty$) returns large values of the KL divergence relative to random (Figure~\ref{si gr:packing dkl}). As a more direct check, we show that the density of neighbors oscillates (Figure~\ref{gr:phase space scan}D), analogous to fairy circle data from reference \cite{tarnitaTheoreticalFoundation2017}. This is not the case for the disordered packing regime, where local repulsion is important but does not lead to long-range order (Figure~\ref{si gr:packing dkl}). Thus, hexagonal packing is confined to a tight region of parameter space in our metabolic growth framework. This region corresponds to a wide separation of timescales, where growth must be sufficiently slow to avoid introducing spatial disorder on the timescales with which relatively fast competitive death stabilizes regular spatial patterning \cite{bonachelaPatchinessDemographic2012,pringleSpatialSelfOrganization2017}.

\section{Transient dynamics and population shockwaves}\label{sec:dynamics}
\begin{figure}[tb]\centering
	\includegraphics[width=\linewidth]{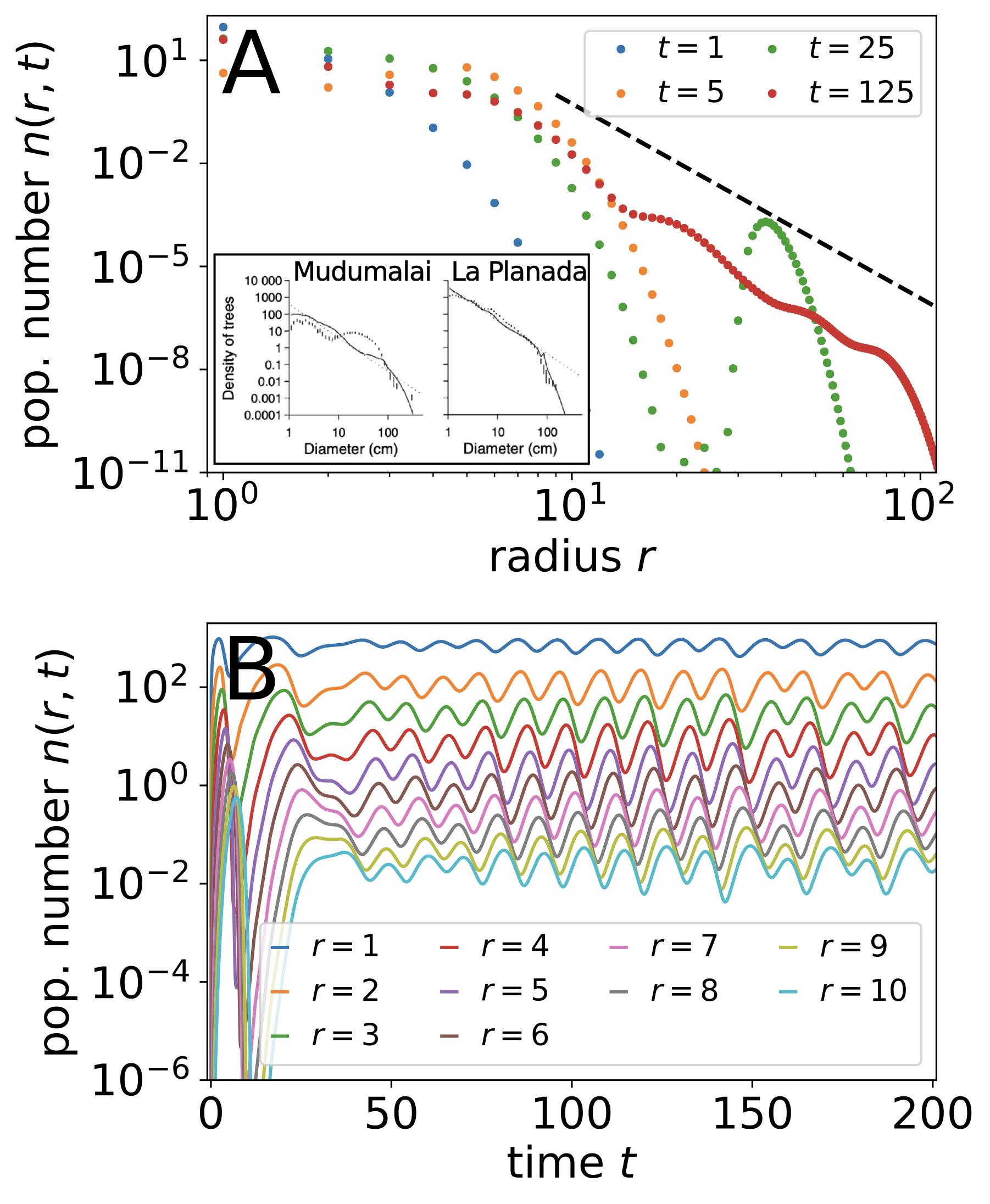}
	\caption{Oscillations in population number $n(r,t)$ from asymmetric competition such as canopy cover. (A) Population number distribution $n(r,t)$ at different times. Insets show data from two tropical forests from reference \cite{muller-landauComparingTropical2006}, where markers correspond to data and lines to their model. Dashed black line shows predicted slope at steady state from Eq~\ref{eq:canopy exponent}. (B) Population number oscilliations for trees of different sizes. See Figure~\ref{gr:2d oscillations} for examples of oscillations in 2D simulation. We use the Heaviside theta function for $1-\Lambda(r'-r)=\Theta(r'-r-\Dr_{\rm crit})$.}\label{gr:oscillations}
\end{figure}

Competitive dynamics may generate harmonic or even chaotic demographic fluctuations as shown by the classic Lotka-Volterra equations describing predator-prey relations \cite{lotkaAnalyticalNote1920,strogatzNonlinearDynamics1994}. Though competitive dynamics are different when organisms are sessile, demographic stability is not guaranteed since ecosystems are buffeted by a wide range of endogenous and exogenous perturbations \cite{levinDisturbancePatch1974,soleSelfsimilarityRain1995,bormannCatastrophicDisturbance1979}. For example, local competition is negligible in a young plot until individuals reach a size where they impinge upon neighbors, a phenomenon known as self-thinning \cite{mradRecoveringMetabolic2020}. This dynamic is especially prominent in agriculture, where spacing is regular, plants are genetically identical, and competition onset is almost uniform \cite{westobySelfThinningRule1984,dengModelsTests2012}. Natural stands also vary with plot age, but they are more stochastic and height differences can be prominent  \cite{enquistExtensionsEvaluations2009,muller-landauTestingMetabolic2006}. Remarkably, in previous measurements we find oscillations in population number with radius as depicted in the inset of Figure~\ref{gr:oscillations}A taken from reference \cite{muller-landauTestingMetabolic2006}, suggesting the presence of long-lived transience not captured by steady state. Inspired by this observation, we consider how competitive asymmetry, specifically forces that decrease fitness of smaller organisms, could excite such population waves.

Asymmetric competition takes various forms such as how canopy shading reduces light incident on shorter plants lying underneath or around the larger ones with little cost to the latter \cite{enquistInvariantScaling2001}. Large termite colonies are much more likely to destroy incipient colonies adjacent to their borders than face a threat \cite{pringleSpatialSelfOrganization2017}. Susceptibility to exogenous disturbances like wind also depends on size though sometimes to the benefit of smaller individuals \cite{everhamForestDamage1996,mitchellWindNatural2013}. As with symmetric competition, we formulate asymmetric competition in terms of its effects on population growth $\dn(r,t)$. A mean-field framework means that the rate of decrease in population number is proportional to typical overlap between trees of radius $r$ with all sizes larger than it up to $\rmax$,
\begin{align}
	n(r,t)\,a_{\rm can}(r) \int_r^{\rmax} n(r',t)\,a_{\rm can}(r')[1-\Lambda(r'-r)] \,dr'.
\end{align}
We assume that the competitive effect $\Lambda(r'-r)$ is some sigmoid-like function that decays from $\Lambda(0)=1$ to the limit $\Lambda(\infty)=0$ (Figure~4 in reference \cite{kempesPredictingMaximum2011}), at which point the tallest trees completely obscure all light incident on ground area spanned by the canopy $a_{\rm can}(r) = c_{\rm can}r^{2\alpha_{\rm can}}$, where $\alpha_{\rm can}=2$ \cite{westGeneralQuantitative2009}. A sigmoidal form indicates some characteristic length scale for $\Lambda$ such that when the distance $r'-r$ reaches some critical value $\Delta r_{\rm crit}$ a substantial portion of light is obscured from above. This is distinct from symmetric interactions that scale with radius $r$ and lack a typical length scale distinguishing competitors from non-competitors. Thus, we consider area-delimited, top-down asymmetric competition which is analogous to canopy shading but more generally captures the competitive advantage of larger organisms \cite{farriorDominanceSuppressed2016,levinCommunityAssembly2001}.

As we show in Figure~\ref{gr:oscillations}B, canopy competition is negligible during initial forest growth but matters strongly when tall trees reach some critical density at which point the difference between the height of the tallest trees and the shortest ones is $\Delta r_{\rm crit}$. At this point, smaller trees are at a sudden competitive disadvantage. Since the maximum tree cutoff is sharp in the population number $n(r,t)$, the appearance of sufficiently large trees for light competition to matter is sudden and causes a correspondingly sharp die-off in young trees, or a population shock. Likewise, the population number of small trees in Figure~\ref{gr:oscillations}A displays a sudden dip at short times. This dip in population number slowly propagates up to larger trees with growth. Eventually, canopy cover dips and small tree population increases suddenly. The delay generates oscillations in population number that may be prominent when competitive interactions are strong and competition with size difference is sharp. Linear stability analysis of the mean-field model suggests that oscillations may be a generic feature of competition between sizes (SI Section~\ref{sec:instability}). In the case of nonlinear metabolic growth, the rightward movement of waveforms in Figure~\ref{gr:oscillations}B accelerates with age: radius grows superlinearly with time $r(t) \propto t^{3/2}$ when metabolic growth exponent $b=1/3$. Besides from superlinear growth, population waves also disperse because of stochasticity in tree growth. Such variation in shape and speed of population shock waves could be used to infer stochasticity in growth and competitive effects following endogenous or exogenous perturbation \cite{levinDisturbancePatch1974,bonachelaPatchinessDemographic2012}.

Remarkably, we find oscillations in population number curves across data on tropical forests. The inset in Figure~\ref{gr:oscillations}A displays two examples from reference \cite{muller-landauComparingTropical2006}. Similar oscillations are visible across other tropical forests. For the example from La Planada, the widths of undulations seem consistent with an example from our simulation in Figure~\ref{gr:oscillations}A though that is not the case for Mudumalai, which shows intriguingly wide oscillations. Forests with repeated prominent demographic oscillations are consistent with long-lived transient oscillations but interestingly that would require tuning of competitive parameters. Yet, another mechanism could be widespread and repeated exogenous perturbations \cite{bormannCatastrophicDisturbance1979}. Overall, competitive dynamics between organisms at different points of maturity can generate oscillatory cycles perhaps influenced by or affecting other classic ecological dynamics \cite{lotkaAnalyticalNote1920,durrettSpatialAspects1998}. Though qualitatively similar curves in demographic data are presented as evidence against metabolic scaling --- indeed space filling yields poor explanation --- our model shows that such deviations may arise due to dynamics overlaying metabolic scaling.

When considering asymmetric competition with resource fluctuations, we find important differences from symmetric competition. Resource constraints impose a limiting cutoff in maximum size and dampen population waves. As in the case of symmetric competition, accounting for resource constraints introduces an exponentially decaying tail that dominates near the point where resource limitations delimit the largest sustainable size. When resource limitations are sufficiently weak that there exists a wide scaling region in the population number that goes as $n(r)\sim r^{-\alpha}$, top-down competition fixes the population number exponent to (see Appendix~\ref{sec:canopy si})
\begin{align}
	\alpha = 4\alpha_{\rm can} +2 - b. \label{eq:canopy exponent}
\end{align}
In contrast with Eq~\ref{eq:alpha}, Eq~\ref{eq:canopy exponent} is free of metabolic growth coefficients but depends on the way that canopy area grows with radius, $2\alpha_{\rm can}$, and metabolic growth exponent $b$. Thus, asymmetric resource competition leads to a different form for scaling exponent $\alpha$ than that of canonical metabolic scaling theory, and its value generally incompatible with $\alpha=2$ because of physical limits on values of $\alpha_{\rm can}$. Though population oscillations likely share exogenous origins, it is remarkable that competition dynamics, though discussed widely in the literature \cite{slatkinModelCompetition1984,westobySelfThinningRule1984}, present one endogenous cause, whose dynamical consequences are hardly remarked upon and suggestively aligned with data.

\section{Discussion}\label{sec:discussion}
The physical structure of a tree is a beautiful fractal not only along its visible constituents, trunk to branches to twigs, but down to the microscopic vasculature that shuttles products of photosynthesis from its self-similar canopy to a branching network of roots. It is remarkable then that even groups of trees seem to obey this pervasive fractal law such that the trees of a particular size ``branch off'' into trees of a smaller size and so on in such a way that we can consider, over some range, the set of large trees as a scaled set of smaller trees \cite{enquistExtensionsEvaluations2009}. This self-similar structure, reflected in power law scaling, emerges from consideration of energetic constraints translated into the requirement that trees fill the available canopy space \cite{westGeneralQuantitative2009}. Yet, other sessile organisms fill space in a variety of ways determined by analogous mechanisms of growth, death, and competition \cite{pringleSpatialSelfOrganization2017}. Inspired by the forest picture, we propose a minimal model of sessile organism growth incorporating aspects of allometric scaling theory and area-based competition. From these basic principles, we obtain a general framework for competitive forces driven by metabolic requirements and fluctuating resources. When interaction with the environment dampens resource fluctuations (e.g.~niche construction) or changes competitive interactions (e.g.~symbiosis), these perturbations will be reflected in the spatial distribution of organisms \cite{pringleSpatialSelfOrganization2017} (Figure~\ref{gr:organism examples}). In this sense, the spatial distribution may serve not only as indicators of changing conditions but also of how competition evolves in altered environments \cite{levinDisturbancePatch1974,tarnitaTheoreticalFoundation2017}. 

We explore such variation by tuning competitive forces in our model with a tractable mean-field theory that succinctly relates metabolic and competitive effects in exponent relations. In the context of resource-area competition, competitive effects are most prominent in the population statistics of the largest organisms. This is because area-delimited competition must scale superlinearly with radius such that it dominates for the largest organisms (Eq~\ref{eq:general mft}). In comparison, individual metabolic growth and mortality scale sublinearly \cite{niklasInvariantScaling2001}, indicating two different regimes of population number for symmetric competition: individual-dominated scaling and competition-mediated cutoffs (Figure~\ref{gr:compartment model2}). Top-down competition, however, exacts a toll in a scale-free way because relatively larger competitors grow, die, and compete the same at every level. Then, competition is manifest in the population number exponent (Eq~\ref{eq:canopy exponent}), affecting both scaling and cutoffs. Our formulation of competitive interactions establishes a basis to be extended to capture environment- or organism-specific variation in resource stress response or sharing. Yet, it also highlights how such diversity converges to universal features summarized by exponents that quantitatively link environmental fluctuations and metabolic scaling (Eq~\ref{eq:stationary soln}).

Besides indicating limitations of metabolic scaling theory --- namely that it may be more accurate in forests with weaker local competition and smaller environmental fluctuations --- probing our theory suggests limitations of spatial-correlation-based measures of regularity when varying organism size introduces disorder in spacing \cite{grohmannMultiscalePattern2010}. As we show by comparing the form of the nearest-neighbor distance distribution with the KL divergence (Eq~\ref{eq:neighbor distribution}), this measure changes weakly with competitive strength, suggesting that statistical approaches to measuring competition are limited. Instead, an integrated framework considering deviations from predicted scaling in demographics as well as spatial patterning may better specify the range of competitive forces acting across environments \cite{adamsSpatialDynamics1995,spainSpatialDistributions1986,tarnitaTheoreticalFoundation2017,levinDisturbancePatch1974,getzinContrastingGlobal2019}.

Beyond competitive forces, we find strong additional constraints necessary to stabilize strong spatial order in models with metabolic growth (Figure~\ref{gr:phase space}). Whereas metabolic scaling tends to inject spatial disorder by constantly changing organism size and by opening free space upon organism death, regular tiling such as seen for fairy circles and some termite mounds requires the elimination of unbridled growth, slow natural mortality, and overwhelming competitive attrition in a background of sparse newcomers. This emergence of order is different, if related to, stripes of vegetation which require some exogenous source breaking rotational symmetry \cite{sherrattAnalysisVegetation2005,sherrattUsingWavelength2015,borgognoMathematicalModels2009,borthagarayVegetationPattern2010}. Thus, hexagonal packing occurs only in a corner of the much broader model space encompassed by our theory (Figure~\ref{gr:phase space}), reflecting the extraordinary nature of such regular patterns. 

Complementary to the connection between spatial patterns and asymmetric competition \cite{levinCommunityAssembly2001,levinsRegionalCoexistence1971}, we explore transient dynamics in the context of a size-based competitive hierarchy \cite{durrettSpatialAspects1998}. Top-down asymmetric competition can couple different time scales to one another and lead to oscillatory modes in population number (Figure~\ref{gr:oscillations}) --- though touching on the topic of self-thinning, our model extends beyond the typical focus on monoculture stands \cite{westobySelfThinningRule1984,slatkinModelCompetition1984,mradRecoveringMetabolic2020,dengModelsTests2012}. When there is a threshold at which such effects become important, such as with canopy light competition, we expect to find similar population shock waves. Remarkably, oscillatory modes manifest in multiple data sets of tropical forest demography \cite{muller-landauTestingMetabolic2006}. Such die-offs also may be observable in other systems or directly measurable if future data collection permits highly resolved temporal data on organism death. Furthermore, the lifetimes of these transient phenomena, indicated by width evolution, may allow us to distinguish internal competitive forces by leveraging demographic perturbation \cite{soleSelfsimilarityRain1995,nobleSpatialPatterns2018}. Oscillatory modes and instabilities are a widely studied feature of biological populations, for example with the classic logistic equation \cite{erneuxAppliedDelay2009,forysDelayedEquations2015}, and metabolic growth in sessile organisms presents an unexplored mechanism by which they arise.

The most striking ecological patterns occur when local interactions generate large-scale regularities, propagating information coherently over large scales and long times \cite{mirolloSynchronizationPulseCoupled1990,nobleEmergentLongrange2015}. Fairy circles and termite mounds are a breathtaking example. Though forests, fairy circles, and termite mounds all seem to obey forces driving the cycle of birth, growth and death at the level of the individual, population-level structure varies widely. Even amongst forests, some are more random like the examples we show here, but others such as the pinyon-juniper ecosystem of the US Southwest are more spatially regular. To connect the wide range of spatial patterns shaped by competitive forces in sessile organisms, we build on theoretical foundations of metabolic scaling. The resulting realm of models may frame analogies between organisms across species, environments, and times in the language of competitive forces churning on top of individual properties constrained by metabolic principles.

\showmatmethods{} 

\acknow{CPK and GBW thank Toby Shannan and CAF Canada for generously supporting this work. GBW and EDL would also like to thank the NSF for their generous support under the grant PHY1838420.}

\showacknow{We acknowledge useful discussions with Sungho Choi about forest data and Deborah Gordon for bringing up self-thinning.} 

\clearpage
\renewcommand{\thefigure}{S\arabic{figure}}
\renewcommand{\theequation}{S\arabic{equation}}
\setcounter{figure}{0}
\setcounter{equation}{0}
\appendix

\section{From discrete to continuum size class model}\label{sec:continuum model}
Though tree population is typically binned into discrete size classes in both observation and theory, tree growth is in reality a function of a continuous radius $r$. In the main text, we show discrete and continuum limits in Eqs~\ref{eq:dnk} and \ref{eq:dn1}. To go from one to the other, we relate the index $k$ to radius $r_k$ such that $r_k \equiv r_0 + k\,\Delta r$. Then, taking Eq~\ref{eq:dnk} and expanding the growth rate in terms of $\Dr$, we obtain $\dr(r+\Dr) \approx \dr(r) + \Dr\partial_r \dr(r) + \mathcal{O}(\Dr^2)$, where the last term contains all terms of quadratic and higher order in minuscule bin width. Likewise, we expand population number $n(r,t)$ about $r$. After rearranging terms, we obtain
\begin{align}
\begin{aligned}
	\dn(r,t) &= -\Dr \partial_r [n(r,t)\dr(r)/\Dr] - n(r,t)\mu(r) + \mathcal{O}(\Dr^2).
\end{aligned}\label{eq:si fp}
\end{align}
In the limit $\Dr\rightarrow0$, we can discard the second and higher order terms $\mathcal{O}(\Delta r^2)$ to find Eq~\ref{eq:dn1}. In Eq~\ref{eq:si fp}, we have made explicit the dependence on bin width, which arises from our definition of growth rate. In this formulation, the number of trees located within the range $[r,r+\Dr)$ that will grow to the next size class in time $dt$ is
\begin{align}
	n(r,t)\dr(r)dt/\Dr,
\end{align}
where we must ensure that the ``distance'' grown $\dr(r)dt$ after a small time step $dt$ is smaller than $\Dr$ to assure that growth rate does not change meaningfully within the bin and that trees do not simply pass through a bin. We must take special care with this limit in the boundary condition for saplings in Eq~\ref{eq:dn2}. When relating coefficients of growth and mortality with their measurements from observational data, such discreteness must be accounted for.

With these equations in hand, we study what happens at steady state by setting the time derivative $\dn(r,t)=0$,
\begin{align}
	0 &= -\dr(r)\partial_r n(r) -n(r)\partial_r \dr(r) - n(r)\mu(r).\\
\intertext{After rearranging terms, we find}
	\frac{\partial_r n(r)}{n(r)} &= \frac{-\partial_r \dr(r) - \mu(r)}{\dr(r)}.
\end{align}
This can be integrated directly after recognizing the left hand size to be the derivative $\partial_r \log n(r)$. Then, the general solution without having specified the metabolic scaling forms for growth and death is
\begin{align}
	n(r) &= n(r_0) \exp\left(-\int_{r_0}^r \frac{\partial_{r'} \dr(r')+\mu(r')}{\dr(r')}\,dr'\right).\label{si eq:steady state soln}
\end{align}
In other words, the steady-state population number depends on the balance of growth rate curvature and mortality with growth rate, determining the total amount of incoming flux as is explicitly solved for the context of metabolic scaling theory in the main text. 

\begin{figure}\centering
	\includegraphics[width=\linewidth]{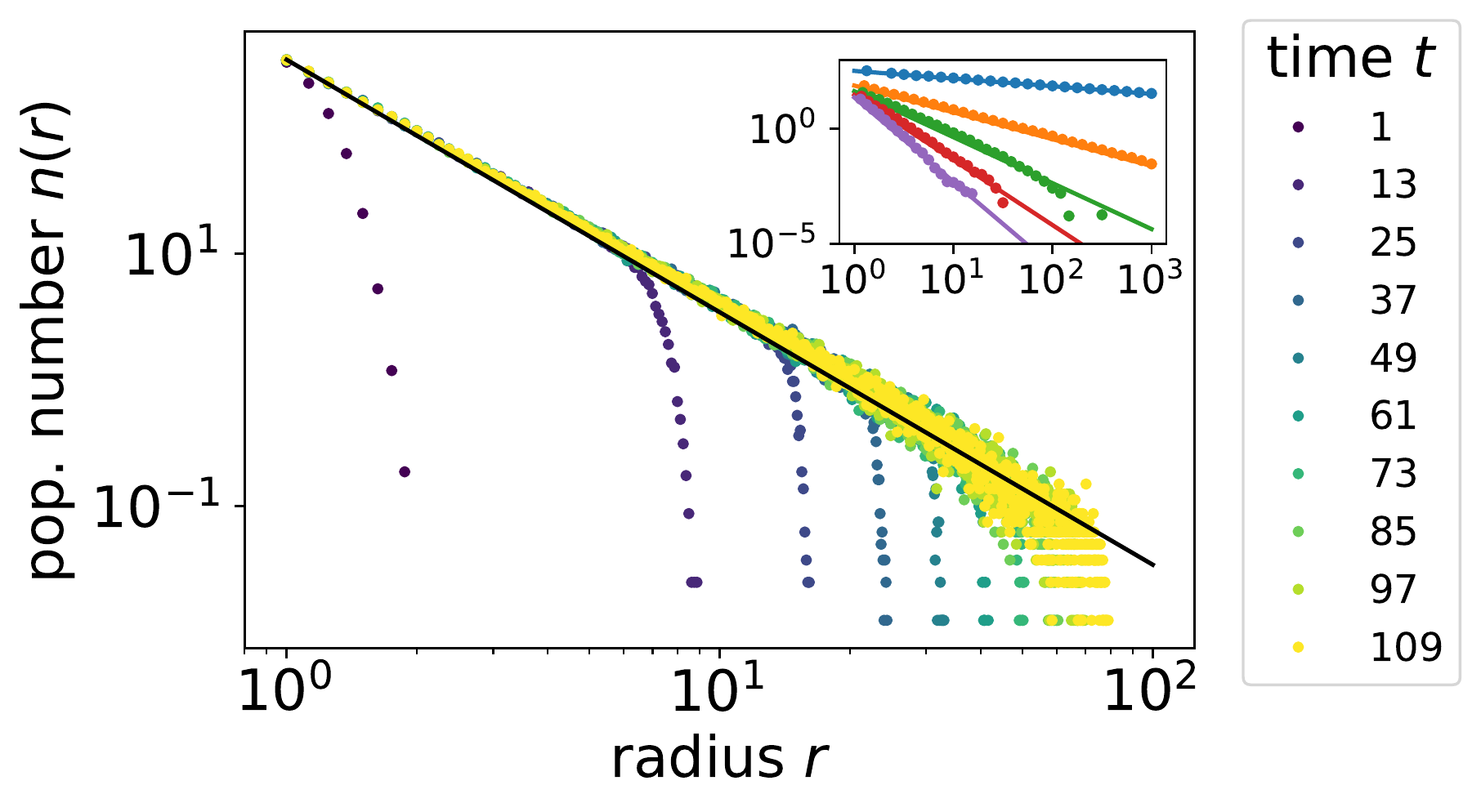}
	\caption{Transience in compartment model starting with empty plot from automaton simulation ($\alpha=2$, $g_0=10^3$ trees per day). (inset) Steady-state profiles for varying scaling exponent $\alpha$ (Eq~\ref{eq:alpha}). Though irrelevant for the continuum limit, discreteness of bins flattens the slope for small radius $r$.}\label{gr:compartment model1}
\end{figure}

Starting with an empty plot, we expect to find transient behavior as individuals grow and fill the available space which must violate steady-state predictions. In the case of the simple metabolic scaling compartment model, the transient is uncomplicated: small trees quickly approach the steady-state scaling form $n(r) \sim r^{-\alpha}$ and a sharp cutoff moves to the right as in Figure~\ref{gr:compartment model1}, where we plot the results of a stochastic automata simulation. When forests are still growing, we would expect such a cutoff to be prominent and even obscure the scaling form. Thus, it is crucial to consider the age of the plot before comparing with steady-state assumptions as is discussed in references \cite{enquistExtensionsEvaluations2009,kempesPredictingMaximum2011,mradRecoveringMetabolic2020}.

\section{Metabolic scaling theory}
In Eq~\ref{eq:allo growth}, we consider an approximation derived from metabolic scaling laws relating radial growth rate in trees with current size measured by radius. This equation derives from the allometric scaling prediction relating biomass growth rate $\dot{m} = \bar{a}m^{3/4}-\bar{b}m$ such that quarter power scaling only dominates for modestly sized organisms, where the exact transition depends on the ratio  of parameters for biological energetics $\bar{a}/\bar{b}$. For interspecies scaling including many examples and a large range, however, data shows the linear term to be negligible, or $\bar{a}/\bar{b}\gg1$ \cite{niklasInvariantScaling2001}. In reference \cite{moriMixedpowerScaling2010}, entire trees are plucked and analyzed to confirm that this quarter-power scaling holds once trees are of $m=0.1$\,kg, which constitute the ``saplings'' we consider. When we use the scaling relation $r=c_m m^{3/8}$ to go from mass to radius, we neglect these corrections and hence the approximation sign in Eq~\ref{eq:allo growth}.

\section{Mean-field theory of symmetric competition}\label{si sec:symmetric competition}
Here, we discuss in deeper detail the derivation of the mean-field theory for symmetric resource competition. In the main text, we focus on the example of tree mortality rate as a function of overlapping root area with neighbors. 

Starting with Eq~\ref{eq:DQ} defining metabolic inequality $\Delta Q$ that must be positive to guarantee survival, we picture placing a tree randomly on the plot and asking with what probability it lands on a region already covered by other trees. Averaging over many spatial arrangements over a long period of time, we can consider the competitive force exerted by others to constitute a kind of ``mean-field,''
\begin{align}
	\DQ &= \rho(t)a(r)\left[ 1 - f\frac{A_{\rm tot}-a(r)}{L^2} \right] - Q_0(r),
\end{align}
where $A_{\rm tot} \equiv \int_{r_0}^{\rmax} n(r') a(r')\,dr'=\rmax^{1+2\alpha_r}/(1-\alpha+2\alpha_r)$ is the total area covered by all individuals for a square plot of linear dimension $L$. In other words, the probability of landing on area occupied by other trees is the simply the fraction of the plot that is covered assuming that trees are not overlapping. This is a key assumption, and in the main text we further assume that the typical ratio of occupied tree area to the plot area is unity, $A_{\rm tot} \approx L^2$, or that all available space is filled. This is clearly a poor approximation when there is much overlap and $A_{\rm tot}>L^2$ or when the plot is sparse such that $A_{\rm tot}<L^2$. However, this dependence ends up only determining the location of the maximum tree size cutoff and not form of the exponential tail given by the exponent $\kappa$.

\begin{figure}
	\includegraphics[width=.8\linewidth]{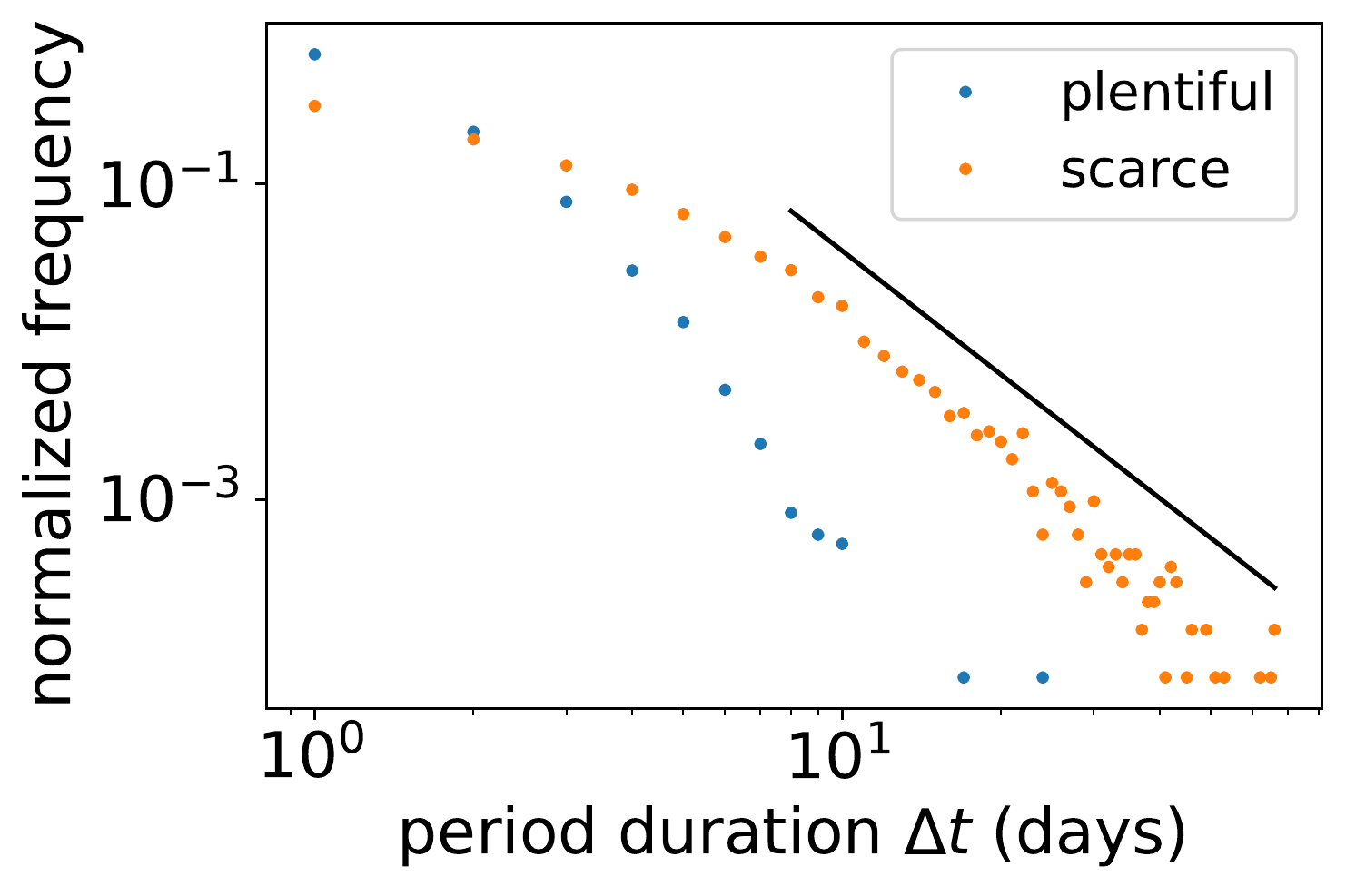}
	\caption{Frequency of periods with above- and below-average rainfall, labeled as ``plentiful'' and ``scarce,'' respectively, in weather stations across Puerto Rico from early 1970s to mid 2010s. To guide the eye, we have included a power law tail $\Delta t^{-2.6}$ as a black line.}\label{gr:si pr rainfall}
\end{figure}

Given this major simplification that leads to Eq~\ref{eq:mft DQ}, the probability of mortality is determined by the probability that incoming resources are not sufficient to cover basal metabolic rate, what we call the ``probability of fatal fluctuations,''
\begin{align}
&\begin{aligned}
	p\left( \xi>\xi_{\rm basal} \right) &= \int_{\xi_{\rm basal}}^\infty h(\xi')\,d\xi'\\
		&=\xi_0^{\nu-1}\int_{\xi_{\rm basal}}^\infty \xi'^{-\nu}\,d\xi'\\
		&= \xi_0^{\nu-1} \frac{1}{1-\nu}\left[ \xi^{1-\nu} \right]_{\xi_{\rm basal}}^\infty\\
\end{aligned}
\intertext{Assuming that $\nu>1$,}
&\begin{aligned}
	p\left( \xi>\xi_{\rm basal} \right) &= \frac{(\xi_{\rm basal}/\xi_0)^{1-\nu}}{\nu-1},
\end{aligned}
\end{align}
where $\xi_0$ is chosen to enforce that the average $\bar\xi=1$. When $\nu\leq1$, the integral diverges with the upper cutoff, and it must be explicitly specified. The upper cutoff must also be specified to calculate a finite mean for $1<\nu<2$, which emphasizes the importance of large fluctuations when $\nu$ is small. Though we primarily refer to the power law scaling form as a reduced representation of the propensity for large resource fluctuations, rainfall has been shown to display large-scale fluctuations suggestive of self-organized criticality \cite{petersCriticalPhenomena2006}. We measure rainfall at NOAA weather stations across Puerto Rico and find that durations of periods showing below average rainfall display a power-law-like tail in Figure~\ref{gr:si pr rainfall} \cite{ClimateData2020}. Nevertheless, we emphasize that the principal role of the power law form is to clearly distinguish between scenarios where large fluctuations are neglible ($\nu>2$) from when they are not ($\nu<2$).

We determine $\xi_{\rm basal}$ by asserting that the metabolic rate inequality in Eq~\ref{eq:mft DQ} is satisfied, leading to
\begin{align}
	\frac{1}{\xi_{\rm basal}} &= \frac{Q_0(r)}{\varepsilon\bar\rho\,a(r)[ 1 - f ]}.
\end{align}
Now, using allometric scaling relations $a(r) = \croot r^{2\aroot}$
\begin{align}
\begin{aligned}
	p(\xi>\xi_{\rm basal}) &= \frac{1}{\nu-1}\left( \frac{\beta_1 \xi_0 r^{\eta_1}}{\varepsilon\bar\rho\,\croot(1-f) r^{2\aroot}} \right)^{\nu-1}\\
	&= \frac{1}{\nu-1}\left( \frac{\beta_1\xi_0}{\varepsilon\bar\rho\,\croot(1-f)} \right)^{\nu-1}r^{(\nu-1)(\eta_1-2\aroot)}
\end{aligned}\label{si eq:mortal fluctuations}
\end{align}
where $\xi_0=(\nu-2)^{1/(2-\nu)}$ to fix the average $\bar\xi=1$ assuming $\nu>2$. In the main text, we recast Eq~\ref{si eq:mortal fluctuations} as Eq~\ref{eq:resource compet} defining exponent $\kappa$ and coefficient $B$. Though the full analytic expression for $B$ is in Eq~\ref{si eq:mortal fluctuations}, we also recognize that the probability of death is assured when organisms surpass basal requirements, or $p(\xi>\xi_{\rm basal})\rightarrow1$. This leads to the expression for $B$ in the main text, which is determined by $\bar\rho$, or typical resource density. 

Then, we solve for the steady state solution by including into Eq~\ref{si eq:steady state soln} an additional term for resource deprivation, $Bs\,r^{\kappa}$, as in Eq~\ref{eq:general mft}. Given $\kappa>b$ and integrating yields
\begin{align}
	\log \left(\frac{n(r)}{n(r_0)}\right) &= -\left(b + \frac{8\bar{A}}{3\bar a c_m^{1-b}}\right)\log\left(\frac{r}{r_0}\right) -\notag \\
		&\qquad\frac{8\,s\,c_m^{b-1}}{3\bar{a}(\kappa+1-b)}B\left(r^{\kappa+1-b}-r_0^{\kappa+1-b}\right).
\end{align}
Exponentiation of both sides gives Eq~\ref{eq:stationary soln}, where the constants have been absorbed into the normalization constant $\tilde n_0$. 

Though symmetric competition does not in principle change metabolic scaling, a sufficiently narrow scaling regime (or correspondingly strong enough of a tail) could mask scaling, effectively nullifying the space-filling assumption of metabolic scaling. In particular, it may be difficult to distinguish metabolic scaling from metabolic-scaling-like tails with limited range of observation such as in reference \cite{westGeneralQuantitative2009}.

\section{Estimating Kullback-Leibler divergence}\label{sec:si dkl}
We use the Kullback-Leibler divergence as a way of comparing the shape of the nearest-neighbor distribution in data or simulation with a random null model. KL divergence represents a fundamental measure of distinguishability between two probability distributions \cite{amariInformationGeometry2016,leeSensitivityCollective2020}, but there are subtleties in estimating information quantities with finite data \cite{leeStatisticalMechanics2015,bialekBiophysicsSearching2012}. Namely, a bin width must be chosen.

\begin{figure}[t]\centering
	\includegraphics[width=\linewidth]{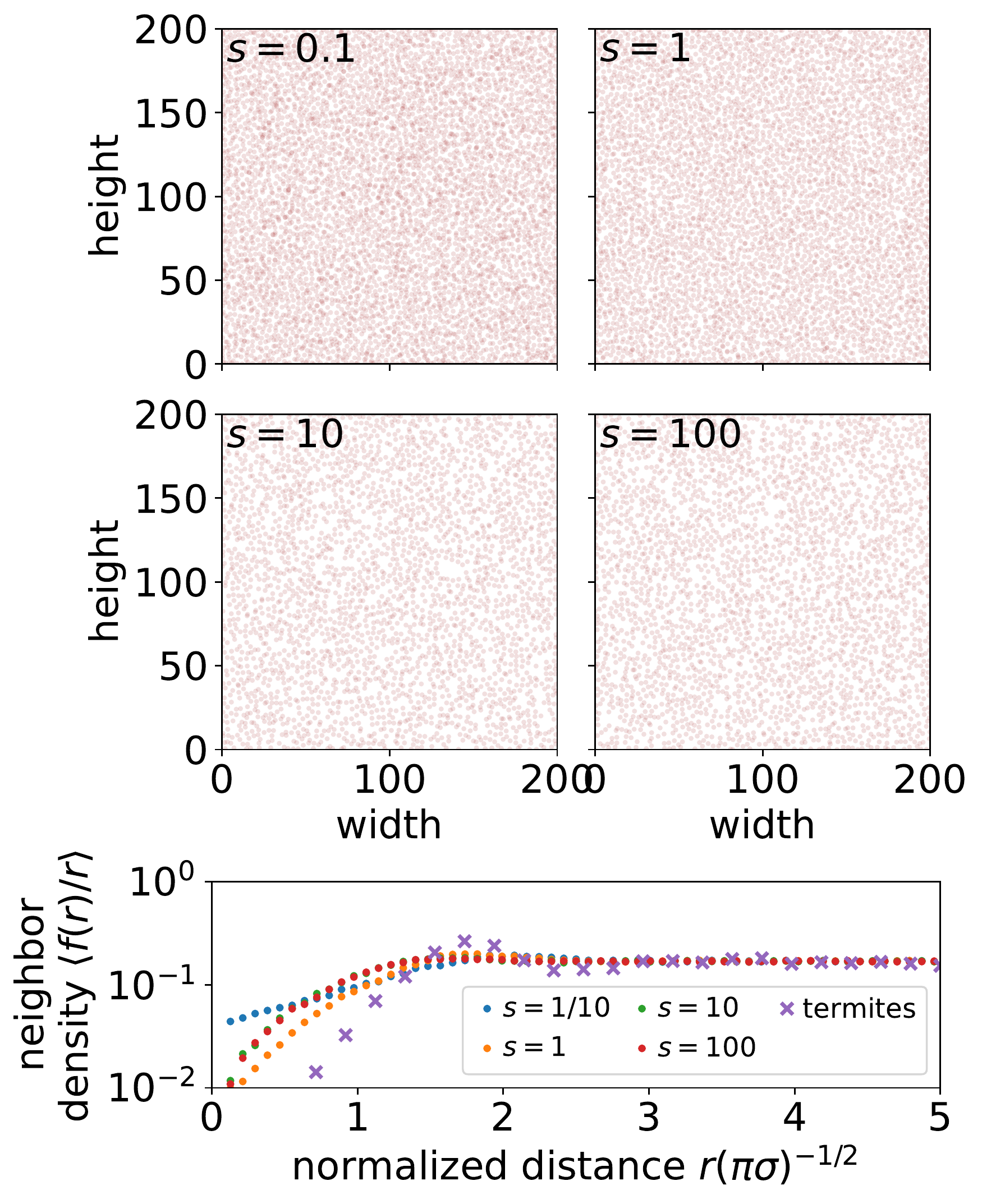}
	\caption{(top) ``Liquid'' phase showing dense packing of individuals without long-range order. These plots are examples of ones on which we measure the KL divergence shown in Figure~\ref{gr:phase space scan}C. When competitive attrition rate $s$ is too large, dense packing is no longer possible, presaging the emergence of roomier hexagonal packing. (bottom) Neighbor density function in disordered packing, or ``liquid,'' phase. Unlike hexagonal packing as in Figure~\ref{gr:phase space}, there is no long range ordering, but local repulsion is evident in the dip near $r=0$.}\label{gr:liquid phase}
\end{figure}

\begin{figure}\centering
	\includegraphics[width=.8\linewidth]{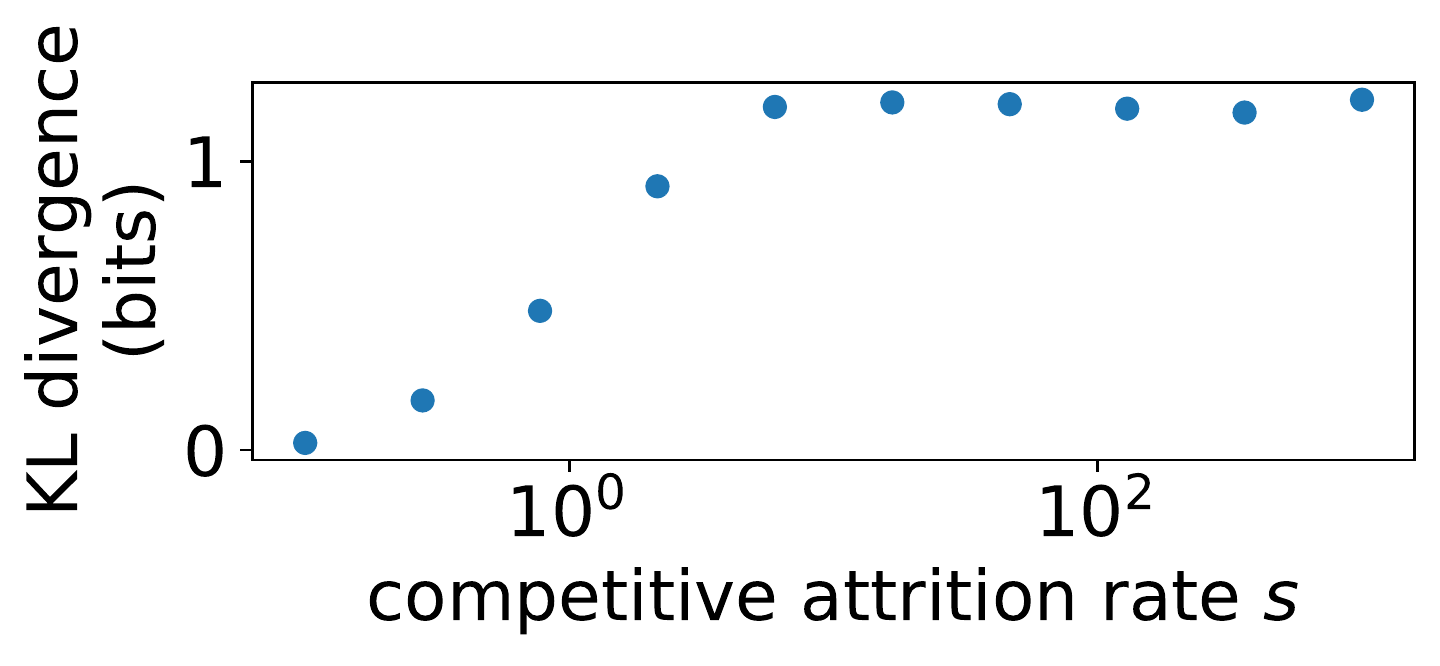}
	\caption{ KL divergence between simulation and random null model distributions of nearest neighbor distance having set $\bar A=0$ and $\bar{a}=0$. Only in this limit does hexagonal packing manifest as is indicated by the large divergence for $s\gg1$. In order to simulate this limit, we set $\bar A=0$ and allow incoming individuals to grow briefly before saturating at a maximum size. This procedure allows us to reach a stable arrangment much faster but corresponds to the same limit when this growth period is brief relative to all other timescales. To see an example of hexagonal packing in a plot of organisms, see Figure~\ref{gr:si termites}.}\label{si gr:packing dkl}
\end{figure}

As defined in the main text and repeated here, the KL divergence between two distributions defined over radius $r$ is
\begin{align}
	D_{\rm KL}(p||q) &= \int_{0}^\infty p(r) \log\left(\frac{p(r)}{q(r)}\right)\,dr.
\end{align}
In the case considered here, we have the analytic form for $q(r)$ in Eq~\ref{eq:dkl}, but we only have a binned approximation to $p(r)$, the estimate $\hat{p}(r) = p(r)\Delta r+\epsilon(r,\Dr, K)$ with error term $\epsilon$ that depends on radius $r$, bin width $\Dr$, and sample size $K$. The corresponding linear approximation of the null distribution is $q(r_{\rm i})\Dr$. Thus, the estimate is
\begin{align}
	D_{\rm KL}(p||q) &\approx \sum_{\rm i} \hat{p}(r_{\rm i}) \log\left(\frac{\hat{p}(r_{\rm i})}{q(r_{\rm i})\Dr}\right)\Delta r,
\end{align}
where the sum is over every unique radius $r_{\rm i}$ in the sample.\footnote{This requires us to assume that $0\log 0=0$, which is justified by continuity \cite{coverElementsInformation2006}.} Using the series expansion $\log(1+x) \approx x - x^2/2 + \mathcal{O}(x^3)$ for $x\ll 1$, we obtain an expansion of the form
\begin{align}
\begin{aligned}
	 D_{\rm KL}(p||q) &\approx \sum_{\rm i} p(r_{\rm i})\left[ \log\left(\frac{p(r_{\rm i})}{q(r_{\rm i})}\right)\Dr + \right.\\
	 &\qquad\left.\frac{\epsilon'}{K} - \frac{\epsilon'^2}{2\Dr K^2} + \mathcal{O}(K^{-3}) \right],
\end{aligned}\label{eq:dkl taylor expansion}
\end{align}
grouping together all terms smaller than order $K^{-3}$ in the last term. When we have a large sample, as we do for our simulations where $K\sim10^5$, then the error terms are determined by finite sampling statistics and so have zero mean but variance that goes like $p(r_{\rm i})\Dr [1-p(r_{\rm i})\Dr]/K$ as has been made explicit here by pulling out the $K$ dependence in $\epsilon$, represented by prime notation $\epsilon'$. Barring the limit of minuscule bin size $\Dr\sim K^{-1}$, where other corrections dominate, we heuristically choose a reasonably small bin size for the large sample set $\Dr=1/20$. Our findings are robust to variation about this choice. Importantly, this allows us to track the emergence of disordered packing, the ``liquid'' phase, and the hexagonally ordered phase as we show in Figures~\ref{gr:phase space scan} and \ref{si gr:packing dkl}, but a more principled analysis of corrections may be required for comparison with smaller data sets.

\section{2D automaton model}
At each time step of duration $dt$, three steps are taken in the following order.
\begin{enumerate}
	\item All tree are grown into the next largest size class with rate given by the growth function $\dot r(r)/\Dr$. Saplings are introduced into uniformly random locations of the plot at rate $g_0$. 
	\item Trees are removed from the system with probability given by mortality rate $\mu(r)$.
	\item If root area competition is included, total resource available to each tree is calculated given a random state of the environment $\xi$. Tree overlap area $\Delta a$ with all neighbors is calculated. Trees falling below the basal metabolic threshold are removed with rate $s\,\Delta a$.
	\item If canopy competition is included, overlapping area $\Delta a$ with taller (more than $\Dr_{\rm crit}$ in height difference) trees is calculated. Trees are moved with rate $s\,\Delta a$.
\end{enumerate}
All variables including set parameters are listed in Table~\ref{tab:parameter list}. As we argue in the main text, the properties of the simulation fall into several generic categories that rescale with the relationships between the considered timescales and dimensions.

\begin{table*}[p]\centering
\caption{Parameters specified in numerical simulations.}\label{tab:parameter list}
\begin{tabular}{c|c}
	parameter 	& 	description \\
	\hline
	$3c_m^{1-b}\bar{a}/8$ &	growth rate coefficient\\
	$a$			& 	root area, symmetric competition area\\
	$\bar a$	& 	biological energetics constant for metabolic growth\\
	$a_{\rm can}$	& canopy area, asymmetric competition area\\
	$A$			&	area of plot\\
	$A_{\rm tot}$	&	total area of all organisms in plot (double-counting overlap)\\
	$\bar A$	& 	natural mortality coefficient\\
	$B$			&	normalization coefficient for $p(\xi>\xi_{\rm basal})$\\
	$b$			& 	metabolic growth rate exponent $\dr \sim r^b$, $b=1/3$\\
	$c_m$		&	coefficient relating mass scaling to radius $r = c_m m^{3/8}$\\
	$c_{\rm r}, c_{\rm h}, c_{\rm can}$	& 	root area (r), height (h), canopy (can) scaling coefficients\\
	$D_{\rm KL}$&	Kullback-Leibler divergence\\
	$F$			&	$8s Bc_m^{b-1}/3$\\
	$f$			& 	resource sharing fraction, $f=1/2$ corresponds to zero-sum game\\
	$g_0$		& 	new organism birth rate\\
	$h(\xi)$	&	probability distribution of scarcity\\
	$K$			& 	number of data points\\
	$L$			&	length of plot\\
	$m$			& 	organism mass\\
	$n(r)$, $\bar{n}(r)$		&	population number at steady state\\
	$n(r,t)$	&	population number as a function of size and time\\
	$p(\xi>\xi_{\rm basal})$	& 	cumulative distribution of scarcity above basal level\\
	$Q$			& 	metabolic rate\\
	$q(r_{\rm min})$	& null random distribution for nearest neighbor distances\\
	$r_0$		& 	smallest organism radius or basal stem radius for trees, i.e.~saplings\\
	$r_k$		&	organism radius for size class $k$\\
	$r_{\rm min}$&	distance to nearest neighbor\\
	$s$			&	death rate when under resource stress\\
	$t$			& 	time\\
	$u(r)$		& 	perturbation function on top of steady state\\
	$\alpha$	& 	population number scaling exponent $n(r) \sim r^{-\alpha}$\\
	$\alpha_1$	& 	basal metabolic rate scaling exponent\\
	$\alpha_{\rm can}$	& canopy radius scaling exponent with radius\\
	$\beta_1$	& 	basal metabolic rate coefficient\\
	$\Delta a$	& 	overlap in area\\
	$\Dr$		& 	size compartment bin width\\
	$\varepsilon$	& 	resource extraction efficiency\\
	$\eta_1$	& 	resource area scaling exponent, $\eta_1=1.8$ for soil water usage\\
	$\kappa$	& 	resource area competition exponent, see $\alpha_1$ and $\nu$\\
	$\Lambda$	&	canopy light permittivity function or lack of larger organism dominance\\
	$\mu_k$		&	natural mortality rate for size class $k$\\
	$\nu$		& 	resource fluctuation exponent\\	
	$\xi$		& 	scarcity\\
	$\rho(t)$	&	resource density\\
	$\bar\rho$	& 	average resource density\\
	$\sigma$	& 	organism density\\
\end{tabular}
\end{table*}

\begin{figure}\centering
	\includegraphics[width=.8\linewidth]{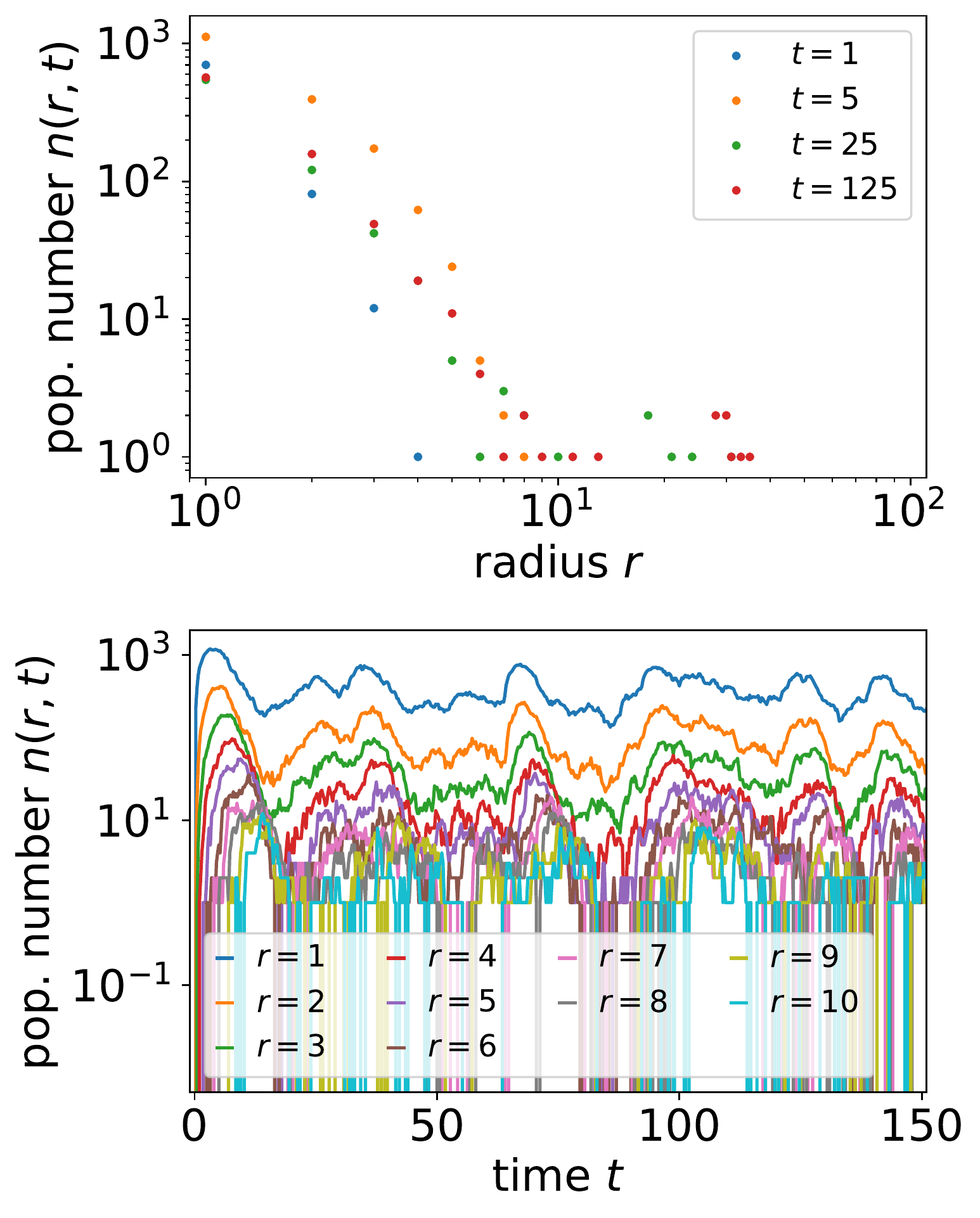}
	\caption{Example of persistent population number oscillations for explicit 2D simulation like in Figure~\ref{gr:oscillations}. For the automaton simulation of individual trees, stochasticity is important.}\label{gr:2d oscillations}
\end{figure}

\section{Canopy light competition exponent}\label{sec:canopy si}
We derive the scaling exponent for population number for resource competition with individuals larger than oneself. Using canopy competition as an example, the mean-field approximation dictates that averaged interaction with trees of the same or of larger size leads to competitive cost
\begin{align}
	n(r)a_{\rm can}(r)\int_r^\infty n(r') a_{\rm can}(r') [1-\Lambda(r'-r)] \,dr',
\end{align}
where the distance function $\Lambda(r'-r)$ indicates how strongly competitive effects come into play with difference in size. We consider some function with a typical length scale $\Dr_{\rm crit}$ which we approximate with the Heaviside theta function
\begin{align}
	n(r)a_{\rm can}(r)\int_r^\infty n(r') a_{\rm can}(r') \Theta(r'-r-\Dr_{\rm crit}) \,dr'.\label{si eq:can1}
\end{align}
The key assumption is of a finite scale at which such effects become important, and the resulting exponent will not depend on whether or not $\Lambda$ is something similar like a decaying exponential or a sigmoid, as is found in forests \cite{kempesPredictingMaximum2011}.

\begin{figure*}\centering
	\includegraphics[width=.8\linewidth]{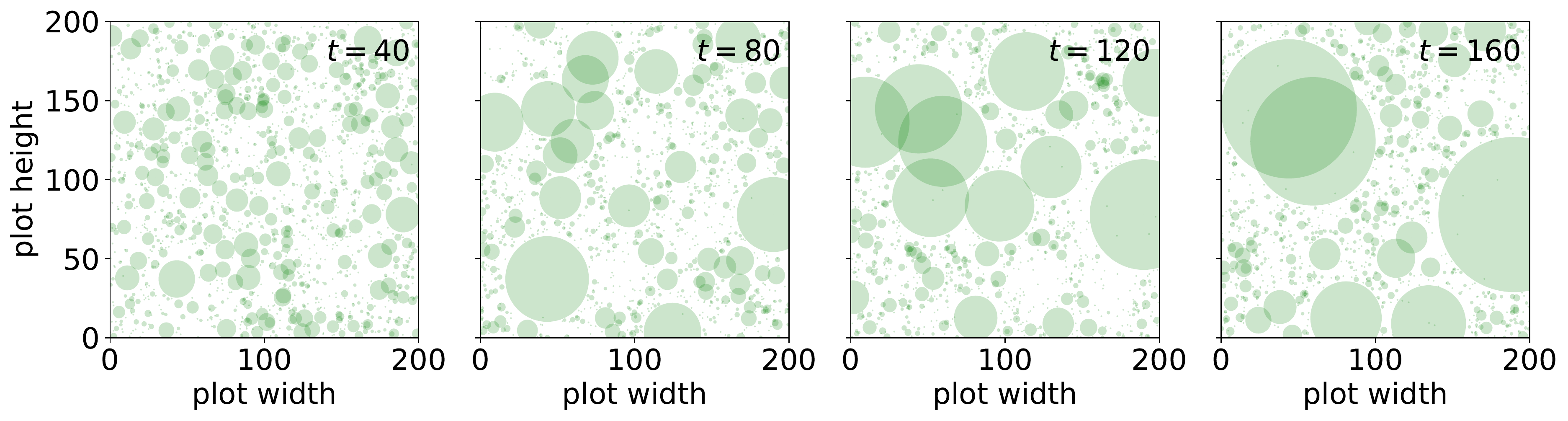}
	\caption{Change in plots over time corresponding to those shown in Figure~\ref{gr:2d oscillations} when starting from an initially empty plot. Green circles show canopy extent.}\label{gr:si oscillation plots}
\end{figure*}

Now, we note that if it is the case that population number $n(r)=c_n r^{-\alpha}$ displays a power law tail and that canopy area scales with radius $r$ as as $a_{\rm can}(r) = c_{\rm can}r^{2\alpha_{\rm can}}$, then the competitive cost is likewise scale-free. Integration of Eq~\ref{si eq:can1} yields
\begin{align}
	=n(r)\frac{c_n c_{\rm can}^2}{\alpha-2\alpha_{\rm can}-1}r^{4\alpha_{\rm can}+1-\alpha}.
\end{align}
This returns us to an analogous form of Eq~\ref{eq:general mft} except now we have asymmetric competition instead of symmetric. At steady state, we have
\begin{align}
\begin{aligned}
	\partial_r [n(r)\dot{r}(r)] &=  - n(r)\mu(r) - n(r) \frac{c_n c_{\rm can}^2}{\alpha-2\alpha_{\rm can}-1} r^{4\alpha_{\rm can}+1-\alpha},\\
	\frac{\partial_r n(r)}{n(r)} &= -\frac{\partial_r\dot{r}(r)}{\dot r(r)} - \frac{\mu(r)}{\dot r(r)} - \frac{c_n c_{\rm can}^2}{\alpha-2\alpha_{\rm can}-1} \frac{r^{4\alpha_{\rm can}+1-\alpha}}{\dot r(r)}.
\end{aligned}\label{si eq:can2}
\end{align}
First, we note that competitive interactions here must decay in a scale-free way for ever larger trees if the population number decays as a power law. This means that the last term in Eq~\ref{si eq:can2} must be commensurate with the others. Symmetric interactions, on the other hand, can be summarized as a constant mean-field effect that scales only with the area of the individual and not with population number. If this is the case, then the scaling exponent must depend on asymmetric competition, implying that the last term goes as $r^{-1}$ and thus we recover Eq~\ref{eq:canopy exponent}. Unlike symmetric resource-area based competition, canopy light competition only depends on the rate of canopy area scaling and the metabolic growth exponent and not the timescales of natural mortality and growth.

In principle, competitive dynamics could replicate WEB metabolic scaling if $\alpha=2$ in Eq~\ref{eq:canopy exponent}. However, this is unlikely because the resource area exponent is lower bounded $\alpha_{\rm can}>1/2$ (otherwise it would not denote an area), which implies that the metabolic scaling exponent $b>2$. Such superlinear growth is clearly at odds with observation which show sublinear scaling in biomass production not just in forests but across diverse biology \cite{niklasInvariantScaling2001,muller-landauTestingMetabolic2006}. This presents a prediction that could be tested by comparing environments where domination by larger organisms displays such scale-free behavior.

\section{Stability analysis}\label{sec:instability}
We analyze the stability of our mean-field theory considering symmetric and asymmetric competition separately and find numerical evidence that instabilities are a generic feature of competitive interactions. This observation aligns with the intuition that population waves propagate through time because of growth and so any particular size population interacts with a delayed version of itself in the future. As is well known in dynamical control theory, population dynamics, and other physical models, self-coupling with time delays generally lead to oscillations and even chaos \cite{erneuxAppliedDelay2009,sahThreeWays2018}.

To analyze the stability of our equations, we linearize them about the steady state solution, denoted by $\bar{n}(r)$. We perturb the steady state with a small correction $\epsilon u(r)$ to obtain
\begin{align}
	\epsilon \dot u(r,t) &= -\partial_r \left(\dr(r) [\bar{n}(r)+\epsilon u(r,t)]\right) - \mu(r)[\bar{n}(r) + \epsilon u(r,t)] - \notag\\
	&\qquad\epsilon\bar n(r) a(r) \int_{r_0}^{r_{\rm max}} u(r',t)a(r')\,dr' - \notag\\
	&\qquad\epsilon u(r,t)a(r)\int_{r_0}^{r_{\rm max}} \bar{n}(r')a(r')\,dr',\label{eq:stab0}
\end{align}
having only kept terms up to linear order of the perturbation $\epsilon\,u(r,t)$ since $\epsilon\ll1$. Though we consider the specific example of symmetric competition in Eq~\ref{eq:stab0}, a similar derivation applies to the asymmetric case.

By equating the terms linear in $\epsilon$, we obtain the following equation for perturbations at each radius $r$,
\begin{align}
	\dot u(r,t) &= -\partial_r [\dr(r) u(r,t)] - \mu(r)u(r,t) - \notag\\
	&\bar n(r) a(r) \int u(r',t)a(r')\,dr' - u(r,t)a(r)\int \bar{n}(r')a(r')\,dr'.\label{eq:stab1}
\end{align}
Eq~\ref{eq:stab1}, because it is linear in time, admits general solutions of the form $u(r,t) = A(r)e^{\lambda(r)t}$, where $\lambda(r)$ determines how perturbations about the steady steady solution behave. This can be determined from the characteristic polynomial of Eq~\ref{eq:stab1}. The real part of $\lambda(r)$ determines if perturbations grow, $\Re[\lambda(r)]>0$, or if they decay, $\Re[\lambda(r)]<0$. When the imaginary component is nonzero, then perturbations will generate oscillations. 

\begin{figure}\centering
	\includegraphics[width=\linewidth]{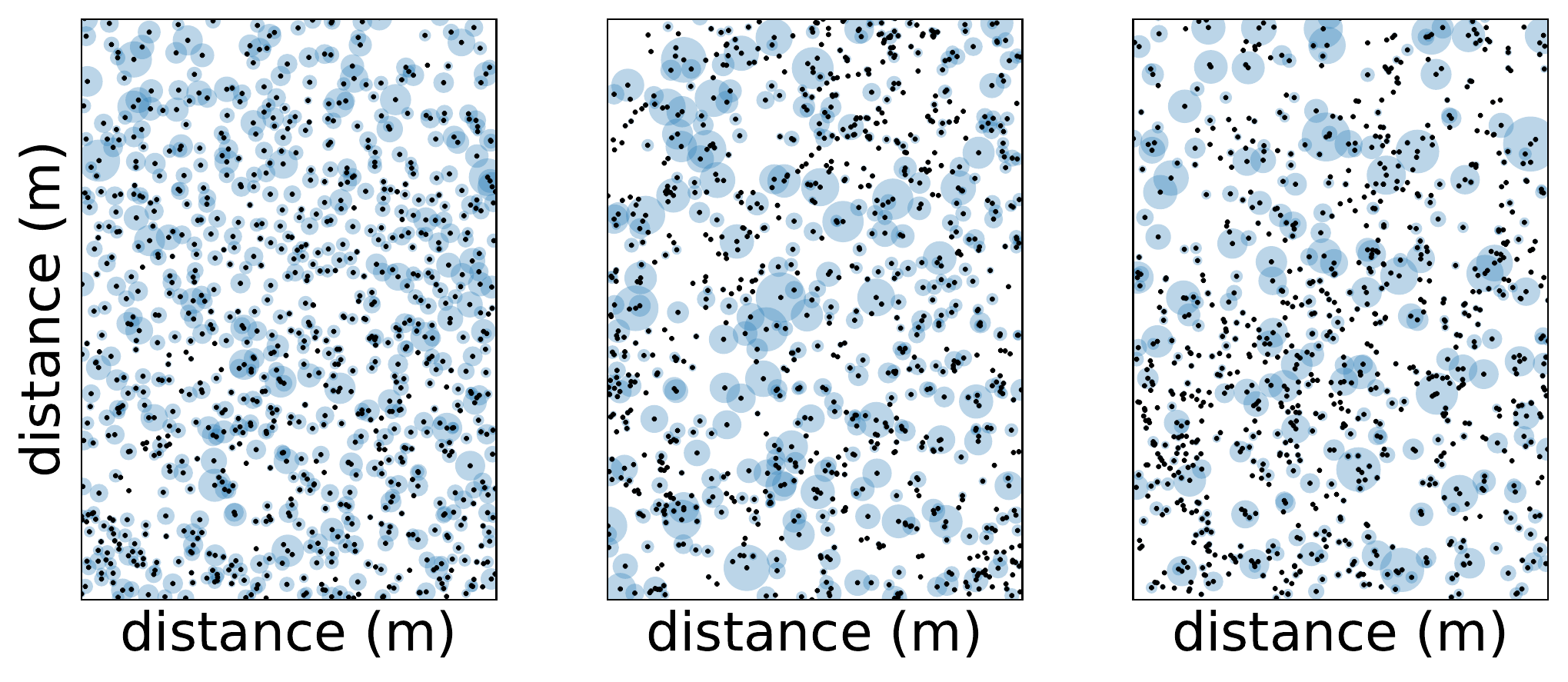}\\
	\caption{Examples of variation in spatial location in Alaskan rainforest plotted using data from reference \cite{schneiderSoutheastAlaska2020}.}\label{gr:si forest examples}
\end{figure}

From numerical calculation of the eigenvalues $\lambda(r)$, we find that typically $\lambda(r)$ is complex when there are competitive interactions, whether interactions are symmetric or asymmetric. Furthermore, we often find that the oscillations are heavily damped though there are some regimes of parameter space showing long persistent oscillations like in Figure~\ref{gr:oscillations}. Interestingly, we also find that there are regimes in phase space where $\Re[\lambda(r)]>0$, indicating unstable directions that could lead to an alternative steady states. Though this mean-field approximation does not completely capture the stochastic 2D simulation, we find similar oscillations as shown in Figure~\ref{gr:2d oscillations}, indicating that population waves are a generic feature arising from delayed self-coupling of populations from metabolic growth.

\section{Survey plot data sets}
In Figure~\ref{gr:si forest examples}, we show additional forest plots from the references cited in the main text for Alaskan rainforests. Though these all constitute similar ecosystems, there is variation between location that may depend on factors like local geography. These effects may be expressed in our parameters for competitive strength and resource fluctuations. 

\begin{figure}\centering
	\includegraphics[width=\linewidth]{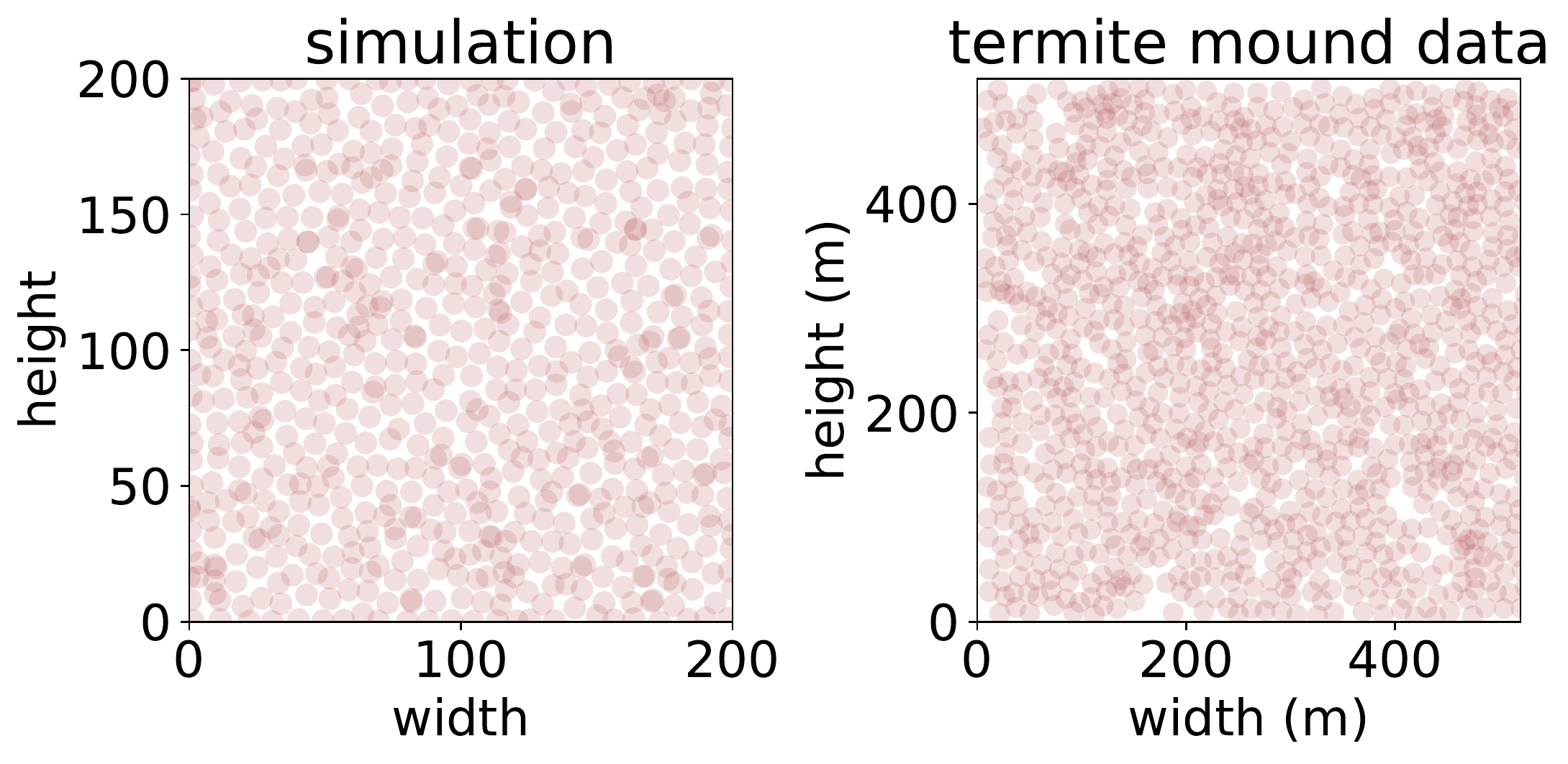}
	\caption{Packed, ``solid'' regime in simulation compared with termite mound packing in Namibia from reference \cite{tarnitaTheoreticalFoundation2017}. See Figure~\ref{gr:phase space scan}D for corresponding spatial correlation functions. As revealed by the neighbor density function there, termite mound packing is tighter than what our simulation recovers by about 10\% (circles indicating termite mounds are meant to be demonstrative and not indicative of actual size or foraging range).}\label{gr:si termites}
\end{figure}

In Figure~\ref{gr:si termites}, we compare the hexagonal packing found in termite mounds with our simulation. 

\bibliography{refs}

\end{document}